\documentclass[journal]{IEEEtran}
\usepackage{amsmath,amsfonts}
\usepackage{algorithmic}
\usepackage{algorithm}
\usepackage{array}
\usepackage[caption=false,font=normalsize,labelfont=sf,textfont=sf]{subfig}
\usepackage{textcomp}
\usepackage{stfloats}
\usepackage{url}
\usepackage{verbatim}
\usepackage{graphicx}
\usepackage{comment}
\usepackage{scalerel}
\usepackage{tikz}
\usetikzlibrary{svg.path}

\usepackage{hyperref} 

\usepackage{array,multirow,makecell}
\usepackage{threeparttable} 
\usepackage{graphicx}
\usepackage{algorithm}

\usepackage{colortbl}
\usepackage{ragged2e}

\usepackage{acronym}

\newacro{vt}[VT]{Virtual Triggering}
\newacro{sca}[SCA]{Side-Channel Attack}
\newacro{scca}[SCCA]{Side-Channel Cryptographic Attack}
\newacro{snr}[SNR]{Signal-to-Noise Ratio}
\newacro{poi}[POI]{Point of Interest}
\newacroplural{poi}[POIs]{Points of Interest}
\newacro{ml}[ML]{Machine Learning}
\newacro{dl}[DL]{Deep Learning}
\newacro{rf}[RF]{Radio Frequency}
\newacro{aes}[AES]{Advanced Encryption Standard}
\newacro{sdr}[SDR]{Software Defined Radio}
\newacro{co}[CO]{Cryptographic Operation}
\newacro{cp}[CP]{Cryptographic Process}
\newacroplural{cp}[CPs]{Cryptographic Processes}
\newacro{ge}[GE]{Guessing Entropy}
\newacroplural{ge}[GEs]{Guessing Entropies}
\newacro{pge}[PGE]{Partial Guessing Entropy}
\newacroplural{pge}[PGEs]{Partial Guessing Entropies}
\newacro{spa}[SPA]{Simple Power Analysis}
\newacro{dpa}[DPA]{Differential Power Analysis}
\newacro{cpa}[CPA]{Correlation Power Analysis}
\newacro{mia}[MIA]{Mutual Information Analysis}
\newacro{em}[EM]{Electromagnetic}
\newacro{fc}[FC]{Frequency Component}
\newacro{pr}[PR]{Pattern Recognition}
\newacro{kr}[KR]{Key Rank}
\newacro{da}[DA]{Differential Analysis} 
\newacro{ca}[CA]{Correlation Analysis} 
\newacro{puf}[PUF]{Physical Unclonable Function}
\newacro{iv}[IV]{Intermediate Value}
\newacro{ro}[RO]{Ring Oscillator}
\newacro{soc}[SoC]{System on Chip}
\newacro{ps}[PS]{Processing System}
\newacro{pl}[PL]{Programmable Logic}
\newacro{nco}[NCO]{Numerically Controlled Oscillator}
\newacro{apu}[APU]{Application Processing Unit}
\newacro{rpu}[RPU]{Real-time Processing Unit}
\newacro{vrm}[VRM]{Voltage Regulator Module}
\newacro{tvla}[TVLA]{Test Vector Leakage Assessment}
\newacro{dac}[DAC]{Digital to Analog Converter}
\newacro{adc}[ADC]{Analog to Digital Converter}
\newacro{des}[DES]{Data Encryption Standard}
\newacro{scca}[SCCA]{Side-Channel Cryptographic Attack}
\newacro{dut}[DUT]{Device Under Test}
\newacro{lsb}[LSB]{Least Significant Bit}
\newacro{ta}[TA]{Template Attacks}
\newacro{pdn}[PDN]{Power Distribution Network}
\newacro{tdc}[TDC]{Time-to-Digital converter}
\newacro{ro}[RO]{Ring Oscillator}
\newacro{psu}[PSU]{Power Supply Unit}
\newacro{slr}[SLR]{Super Logic Region}
\newacro{ip}[IP]{Intellectual Property}
\newacroplural{ip}[IPs]{Intellectual Properties}
\newacro{fpga}[FPGA]{Field-Programmable Gate Array}
\newacro{iot}[IoT]{Internet of Things}
\newacro{asic}[ASIC]{Application-Specific Integrated Circuit}
\newacro{scare}[SCARE]{Side-Channel Analysis for Reverse Engineering}

\usepackage{orcidlink}

\begin{document}

\title{Multi-Screaming-Channel Attacks: Frequency Diversity for Enhanced Attacks}

\author{
Jeremy Guillaume $^{\orcidlink{0009-0005-3398-3423}}$\,, \and 
Maxime Pelcat $^{\orcidlink{0000-0002-1158-0915}}$\,, \and
Amor Nafkha $^{\orcidlink{0000-0002-1164-7163}}$\,, \and
and Rubén Salvador$^{\orcidlink{0000-0002-0021-5808}}$,\,\IEEEmembership{Member,~IEEE}

\thanks{Manuscript submitted on April 3, 2025.}
\thanks{Jeremy Guillaume is with Université Bretagne Sud, CNRS, Lab-STICC, Lorient, France (email: jeremy.guillaume@univ-ubs.fr).}
\thanks{Amor Nafkha is with CentraleSupelec Rennes, CNRS, IETR, France (email: amor.nafkha@centralesupelec.fr).}
\thanks{Maxime Pelcat is with INSA Rennes, CNRS, IETR, France (email: maxime.pelcat@insa-rennes.fr).}
\thanks{Ruben Salvador is with CentraleSupelec, Inria, CNRS, IRISA, Rennes, France (email: ruben.salvador@inria.fr).}}

\markboth{Journal of \LaTeX\ Class Files,~Vol.~14, No.~8, August~2015}%
{Shell \MakeLowercase{\textit{et al.}}: A Sample Article Using IEEEtran.cls for IEEE Journals}

\maketitle

\begin{abstract}
Side-channel attacks consist of retrieving internal data from a victim system by analyzing its leakage, which usually requires proximity to the victim in the range of a few millimetres.
Screaming channels are EM side channels transmitted at a distance of a few meters. 
They appear on mixed-signal devices integrating an RF module on the same silicon die as the digital part.
Consequently, the side channels are modulated by legitimate RF signal carriers and appear at the harmonics of the digital clock frequency.
While initial works have only considered collecting leakage at these harmonics, late work has demonstrated that the leakage is also present at frequencies other than these harmonics.
This result significantly increases the number of available frequencies to perform a screaming-channel attack, which can be convenient in an environment where multiple harmonics are polluted.
This work studies how this diversity of frequencies carrying leakage can be used to improve attack performance.
We first study how to combine multiple frequencies.
Second, we demonstrate that frequency combination can improve attack performance and evaluate this improvement according to the performance of the combined frequencies.
Finally, we demonstrate the interest of frequency combination in attacks at $15$ and, for the first time to the best of our knowledge, at $30$ meters. 
One last important observation is that this frequency combination divides by $2$ the number of traces needed to reach a given attack performance.
\end{abstract}


\IEEEpeerreviewmaketitle

\section{Introduction}
\IEEEPARstart{S}{ide-channel attacks}~\cite{choi2020tempest, standaert2010introduction} allow attackers to retrieve confidential information from computing devices by exploiting the correlation of internal data with the leakage produced while computing over these data.
The term \emph{side channel} is therefore used to refer to physical leakage signals carrying confidential information. 
Side channels are general to CMOS computing devices and can take many forms, from runtime variations of system power consumption~\cite{mangard2008power} to \acf{em} emanation\cite{gandolfi2001electromagnetic}. 
Screaming channels~\cite{camurati2018screaming} are a specific form of \ac{em} side channel that occurs on mixed-signal devices, where a \ac{rf} module is co-located on the same die as digital modules. 
In this context, the leakage of the digital part reaches the \ac{rf} module, which can transmit it over a distance of several meters. 
This phenomenon allows attackers to mount side-channel attacks at larger distances from the victim than regular side-channel attacks, which traditionally require very close access to targets. 
The seminal work from Camurati et al.~\cite{camurati2018screaming} demonstrated how screaming-channel attacks can succeed at distances of up to 15 meters. 

Leakage, generated by the switching activity of the transistors from the digital part of the victim system, operates at a clock frequency $F_{clk}$. 
When observed on a spectrum analyzer, the leakage power spectral density is shaped as peaks at the harmonics of $F_{clk}$ ($i.e.$ $n\times F_{clk}$ where $n \in  \mathbb{Z}$). 
What makes screaming-channel attacks different from other \acp{sca} is that the harmonics, after being modulated by the \ac{rf} module, are visible around the carrier frequency $F_{RF}$ of the legitimate \ac{rf} signal. 
Contrary to previous works, Guillaume et al. demonstrated that exploitable leakage is also present at non-harmonic frequencies~\cite{guillaume2023attacking}.
This increases the number of potential frequencies from which an attacker can select to build successful screaming-channel attacks.

Previous studies on side-channel attacks demonstrate that combining multiple channels, \emph{i.e.}, multiple sources of leakage, can improve the attack performance~\cite{agrawal2003multi, standaert2008using, elaabid2011combined, hutter2012exploiting, souissi2012towards, heyszl2014clustering, specht2015improving, yang2017multi, genevey2019combining, yang2023mca, ayoub2025phasesca}. 
Indeed, the different channels can carry complementary information on the target data or be affected by statistically independent noise.

In this paper, \textbf{we study how combining the leakage carried by different frequencies can improve screaming-channels attacks performance}.
We believe the different frequencies should carry common information on data computed internally by the victim device but should differ in the noise as they are spaced enough to have independent noises.
We \textbf{propose considering these frequencies as different channels in a multi-channel attack} and answer the following questions.
\begin{itemize}
    \item \textbf{How to combine multiple frequencies, \emph{i.e.}, their carried leakages, for screaming-channel attacks?}
    \item \textbf{What impact does frequency combination have on the performance of screaming-channel attacks?}
\end{itemize}

\emph{This work brings state-of-the-art multi-channel attack methods that combine multiple side channels into the particular case of screaming channels, interpreting the different frequencies as individual leakage channels}. Our contributions are:
\begin{itemize}
    \item We propose \textbf{multi-screaming-channel attacks} by  considering \textbf{different frequencies as separate side channels} in Section~\ref{sec:multi-channels_in_screaming_channels}, and we \textbf{compare the performance of the different combination methods} to find which performs best for screaming-channel attacks. 
    \item We demonstrate the interest of \textbf{frequency combination to improve attack performance} in Section~\ref{sec:att_improvement} in two experiments in laboratory conditions:
    \begin{itemize}
        \item We show how to \textbf{mount successful attacks with fewer traces and analyze the impact of the frequency diversity order} in Section~\ref{sec:weak_freqs}. This scenario considers attackers that can only collect a limited number of traces, with which all frequencies are too weak for a successful attack.
        \item We \textbf{quantify the improvement in attack performance} according to the initial performance of the combined frequencies in Section~\ref{sec:combination_efficiency}. 
    \end{itemize}
    \item We prove the interest of \textbf{frequency combination in realistic scenarios}, reducing the number of needed traces and increasing the attack distance in Section~\ref{sec:realistic_attacks}:
    \begin{itemize}
        \item We mount a successful \textbf{attack at $15$~meters with fewer traces} and discuss how the results from our approach compare to current art in Section~\ref{sec:attack_15m}.
        \item We mount the \textbf{first successful attack at the largest distance reported, $30$~meters}, in Section~\ref{sec:attack_30m}.
    \end{itemize}
\end{itemize}

The rest of the paper is organized as follows: Section~\ref{sec:soa} reviews the background and presents the related work. Section~\ref{sec:multi-channels_in_screaming_channels} describes how to repurpose multi-channel attacks as \emph{multi-screaming-channel attacks}, while Section~\ref{sec:att_improvement} proves their interest and quantifies the improvement brought over base attack performance using only one single frequency. Drawing from these observations, in Section~\ref{sec:realistic_attacks} we report more realistic attacks, comparing our work with current art attacking at $15$~meters, and reporting the first successful attack at $30$~meters. Finally, we discuss our results in Section~\ref{sec:discusion} and conclude the paper in Section~\ref{sec:conclusion}.

\section{Related works on multi-channel attacks}
\label{sec:soa}
In the context of side-channel attacks, a multi-channel attack combines multiple sources of leakage.
These attacks are introduced by Agrawal et al.~\cite{agrawal2003multi}.
Authors propose combining traces from an \ac{em} and a power channel to attack \acf{des}.
Combining multiple channels can increase the quantity of collected information on the targeted data or reduce noise by exploiting the noise independence between the combined channels.
This is expected to increase the \ac{snr} and, thus, the attack performance.

\begin{figure*}[tb]
\centerline{\includegraphics[width=0.8\textwidth]{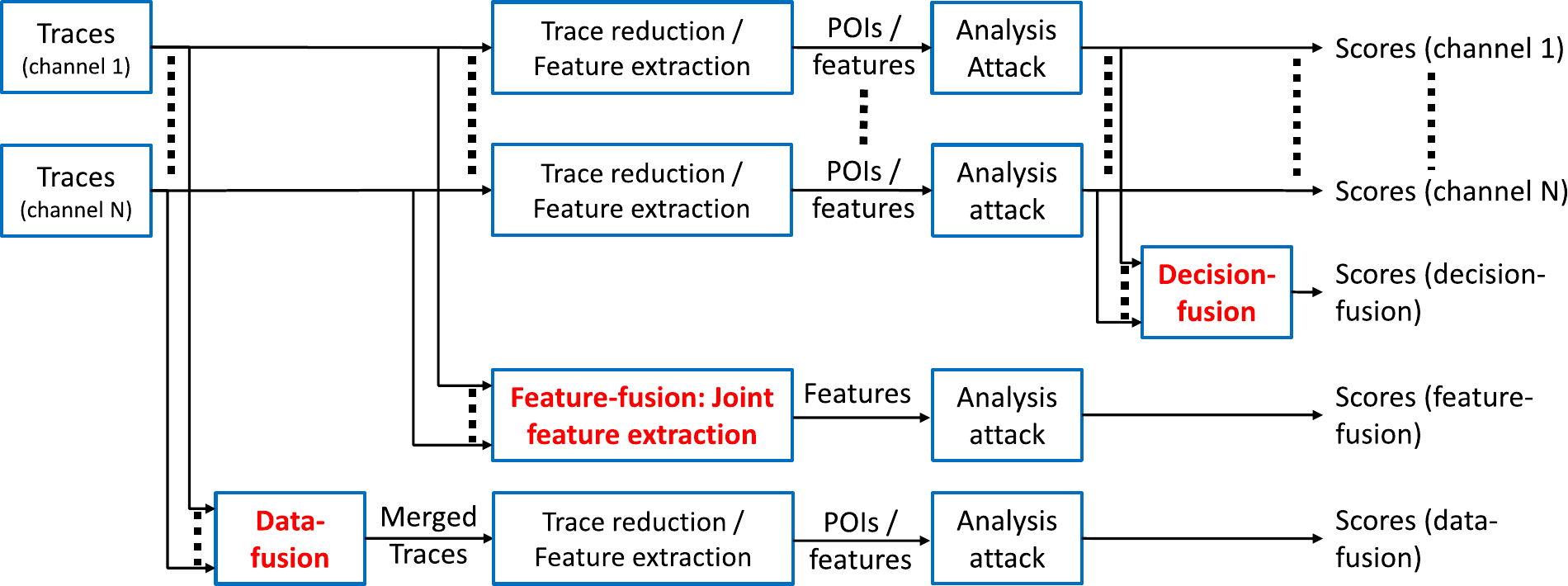}}
\caption{\textbf{State-of-the-art on combination methods:} Leakage from $N$ channels can be combined using $3$ different strategies. 
1/ Data fusion: merges the trace data from the different channels directly.
2/ Feature fusion: extracts features from data of all channels jointly.
3/ Decision fusion: merges decisions from attacks performed independently on each channel.
POIs: Points of interest
}
\label{fig:c5:combination_methods}
\end{figure*}

In this section, we discuss the existing methods to combine multiple channels.
We use the classification proposed by Yang et al.~\cite{yang2023mca} to categorize the combination methods in $3$ groups: \emph{data fusion}, \emph{feature fusion}, and \emph{decision fusion}.
These three combination strategies are illustrated in Fig.~\ref{fig:c5:combination_methods}.
Initially, leakage traces are collected from $N$~channels corresponding to any side channel vector like power, \ac{em}, timing, etc.
In the attack chain, \emph{trace reduction} is a step that can be applied before the analysis attack.
Reducing the trace dimensions reduces the computational complexity during the analysis attack since it reduces the number of samples to analyze.
This reduction can be done by identifying the \acp{poi}, to keep only these trace samples, or by applying \emph{feature extraction} methods like the well-known Principal Component Analysis (PCA)~\cite{kurita2019principal} on the traces.
These methods aim to reduce trace dimensions while keeping the components of the traces that maximize the remaining information on the target data after trace reduction.
Finally, the \emph{analysis attack} analyzes the leakage traces and returns scores, \emph{i.e.}, a probability for each key hypothesis to correspond to the correct key.
In the following, we discuss how the three combination strategies are applied in previous works to combine these $N$~attack chains and obtain a multi-channel attack.

\paragraph{Data-fusion} It consists of directly merging the leakage traces collected from the victim over the same operations.
For example, Hutter et al.~\cite{hutter2012exploiting} collect power traces from two identical devices that compute the same operations. 
Authors average their traces, which also averages their noise, thus reducing it. 
Another way to perform data fusion is by placing traces collected from the different channels in a common trace dataset. 
For example, Heyszl et al.~\cite{heyszl2014clustering} collect \ac{em} leakage of an FPGA using $9$~probes placed at different positions.
They demonstrate that the attack using the traces from all positions performs better than the best of the individual attacks at each position.
Genevey et al.~\cite{genevey2019combining} concatenate traces from the different channels and perform analysis attacks based on machine-learning approaches.
They observe different improvements in attack performance depending on the contribution of the additional channel; in some cases, the channels can add more noise than additional information.

\paragraph{Feature-fusion} It can be considered as a particular case of data fusion.
Compared to data fusion, feature fusion does not directly merge traces together. 
Instead, it jointly extracts features from traces of the multiple channels using feature extraction methods~\cite{standaert2008using, specht2015improving, specht2018dividing, hettwer2020deep}. 
Feature fusion is proposed by Standaert et al.~\cite{standaert2008using} where they apply feature extraction methods on concatenated traces from two channels, an \ac{em} and a power channel. 
Authors compare Principal Component Analysis (PCA)~\cite{kurita2019principal} and Fisher’s Linear Discriminant Analysis (LDA)~\cite{xanthopoulos2013linear} as feature extraction methods and observe that LDA extracts information better than PCA.

\paragraph{Decision-fusion} It combines results from mono-channel attacks performed separately for each channel. 
For example, an analysis attack against \ac{aes} returns $256$ probabilities, \emph{i.e.}, scores, for each \ac{aes} sub-key.
The highest probability is expected to correspond to the correct sub-key value.
In multi-channel attacks at the decision level, the $256$ probability values from the different attacks are merged using an aggregation function, \emph{e.g.}, summing or averaging these probabilities.
Souissi et al.~\cite{souissi2012towards} studied the use of these two aggregate functions, \texttt{sum()} and \texttt{max()}, and observed that \texttt{sum()} performs slightly better than \texttt{max()} function in their context.

In the context of screaming-channel attacks, Camurati et al.~\cite{camurati2020understanding} investigate spatial diversity, \emph{i.e.}, the combination of traces collected at the same frequency with two antennas spaced by $3$~cm.
The combination is performed with the data fusion strategy, by averaging the traces together.
Before this, authors implement a method to equalize channels.
To this end, they compare Equal Gain (EG) and Maximal Ratio Combination (MRC) methods and observe that MRC is better than EG.
In an attack at $0.55$~meter from the victim, the spatial diversity solves the difficulty when attacking in non-Line-of-Sight (nLoS) conditions, \emph{i.e.}, in an environment with obstacles between the attacker and the victim.

\section{Multi-screaming channel attacks: Multi-channel attacks in the context of screaming channels}
\label{sec:multi-channels_in_screaming_channels}

Previous work~\cite{guillaume2023attacking} in screaming channels proved how compromising leakage signals could be found and exploited not only at the harmonics of the digital clock frequency of the victim but also widely spread throughout the EM spectrum at non-harmonics.
As a result, the authors demonstrated how to mount successful screaming-channel attacks in cases where traditionally used frequencies were highly polluted due to other colliding signals like WiFi.

In this work, we \emph{leverage this fact of leakage being present and exploitable in several locations of the EM spectrum} and propose \textbf{multi-screaming channel attacks}, which we formulate as screaming-channel attacks that leverage leakage captured at different frequencies by exploiting combination methods from multi-channel attacks. 

In this Section, we evaluate these combination methods when the different channels are considered to be leakage signals captured at different frequencies.
Our goal is to find the combination method that fits the best for multi-screaming channel attacks before formally assessing the impact of the combination on attack performance in Section~\ref{sec:att_improvement}.

\subsection{Experimental setup and attack evaluation}
To perform our experiments, we collect trace sets at $150$~frequencies (from $2.450$~GHz to $2.6$~GHz, spaced by $1$~MHz).
The setup is the same as the one used by Guillaume et al.~\cite{guillaume2023attacking}, 
with the victim being a PCA10040\footnote{\url{https://www.nordicsemi.com/Software-and-tools/Development-Kits/nRF52-DK.}}, that integrates a nRF52832\footnote{\url{https://www.nordicsemi.com/Products/nRF52832}}, a \acf{soc} from Nordic Semiconductor, and a USRP N210\footnote{\url{https://www.ettus.com/all-products/un210-kit/}} as the radio used by the attacker to collect victim's leakage.
To simplify the evaluation of the combination methods and the improvement in attack performance, we consider ideal laboratory conditions for our experiments in this and in Section~\ref{sec:att_improvement}.
For this, we collect traces by cable, which avoids environmental pollution and increases the number of exploitable frequencies that can be combined.
In section~\ref{sec:realistic_attacks}, to study the impact and build attacks in more realistic conditions, traces are collected at a distance using a directive antenna with a gain of $26$~dBi.
Traces are collected with a time diversity order of $10$, \emph{i.e.}, 
multiple traces collected from the same leakage source and from \acp{cp} that compute the same data, \emph{i.e.}, the same plaintext and key, are averaged together. 
Therefore, under the assumption that the noise has a Gaussian distribution, averaging $N_{TimeDiv}$ segments of noise divides the noise energy by $\sqrt{N_{TimeDiv}}$.

The target victim process runs tinyAES \ac{aes}-128 compiled without optimizations\footnote{tinyAES implementation included in Nordic Semiconductor SDK: \url{https://www.nordicsemi.com/eng/Products/Bluetooth-low-energy/nRF5-SDK}.} and running on one instance of the nRF52832.
To recover the secret key, the trace reduction phase involves selecting \acp{poi} containing information on \acp{iv}.
Then, we perform the profiled correlation attack~\cite{durvaux2016improved}.
Most of the time, the attack does not directly recover the correct key, \emph{i.e.}, the \ac{kr}, which corresponds to the rank of the correct key among all the possibilities in decreasing order of their probability is higher than $0$.
In case this \ac{kr} is low enough, a brute-force attack~\cite{poussier2016simple} can test the ranked possibilities and recover the correct key in a reasonable time.

\ac{kr} is a convenient metric to evaluate attack performance in contexts where the attacks do not always recover the full key.
As in previous works on screaming channels~\cite{camurati2020understanding}, we use \ac{ge}~\cite{kopf2007information} 
which is computed from the \ac{kr} as expressed in Eq.~\eqref{eq_GE} to return concise attack results.
When the \ac{ge} equals $32$, our experimental computer (4-core Intel Xeon(R) CPU E3-1226 V3 @ 3.30 GHz and 8 GB RAM) takes about $5$~minutes to brute-force the key. 
When it equals $35$, the brute-force attack takes about $1$~hour, and approximately $1$~day when it equals $39$.
When performing $N$~attacks in the same conditions to ensure statistically significant results, the attack score will be the average of the $N$~\acp{ge} as expressed in Eq.~\eqref{eq_avg_GE}.

\begin{equation}
\label{eq_GE}
GE = Log_{2}(\ac{kr}).
\end{equation}

\begin{equation}
\label{eq_avg_GE}
GE = \frac{1}{N} \sum_{i=1}^{N} GE_{i}.
\end{equation}

\subsection{Comparing the combination strategies}
\label{sec:combination_strategies_screaming_channels}
Starting from the initial attack chain used in previous works, we propose to test and compare the data fusion and decision fusion methods and observe which method fits the best in the context of screaming-channel attacks.
The study of feature fusion is kept for future work since we do not use a feature extraction method like PCA or LDA.
To simplify the analysis, this first study considers a situation where only two frequencies, $f1$ and $f2$, are combined.

\subsubsection{Data fusion}
\label{sec:c5:data_fusion}
\begin{figure}[tb]
\centering
\subfloat[]{\includegraphics[width=0.16\textwidth]{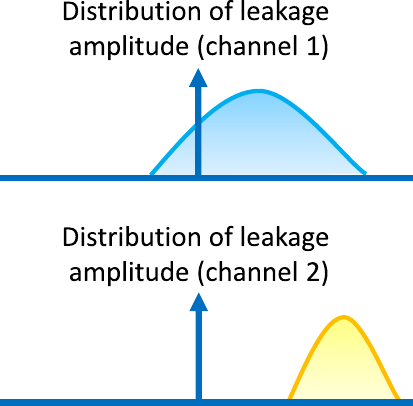}
\label{fig:c5:data_fusion_step_1}}
\subfloat[]{\includegraphics[width=0.16\textwidth]{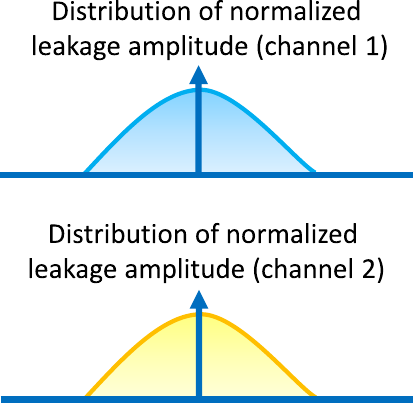}
\label{fig:c5:data_fusion_step_2}}
\subfloat[]{\includegraphics[width=0.16\textwidth]{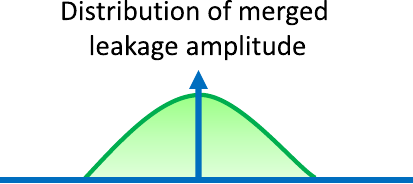}
\label{fig:c5:data_fusion_step_3}}
\caption{\textbf{Data-fusion steps:}  
(a) Leakage amplitude of the two initial channels at samples corresponding to \acp{poi}. 
(b) Normalized values with equivalent amplitudes between the two channels.
(c) Merged values: averaging the normalized values.
}
\label{fig:c5:data_fusion_steps}
\end{figure}

The hypothesis under a data fusion strategy is that leakage is expected to have similar amplitude over the two channels but with independent Gaussian distribution noise~\cite{hutter2012exploiting, heyszl2014clustering}.
Therefore, averaging leakage from these two channels also averages this noise, which, as with time diversity, improves the \ac{snr}.
However, in the context of screaming channels, the two channels correspond to two frequencies, $f1$ and $f2$, in the radio frequency range where the digital leakage is transmitted.
Leakage can undergo different attenuation and distortions through the path between transmitter and receiver from one frequency to the other.
Therefore, the leakage can have different amplitudes in the two channels and must be equalized before being combined (Fig.~\ref{fig:c5:data_fusion_step_1}).
\emph{Pre-processing} steps are added to achieve this equalization. 
First, raw leakage amplitudes are normalized for every channel using z-score normalization (Fig.~\ref{fig:c5:data_fusion_step_2}).
Thus, the leakage amplitudes from both channels will have the same amplitude range and can be merged.
The traces are averaged together, returning a merged trace as if it came from a single channel (Fig.~\ref{fig:c5:data_fusion_step_3}).
These merging steps are performed only after the \acp{poi} selection, \emph{i.e.}, the trace reduction phase.
This reduces the computation complexity since these merging steps are performed only on \acp{poi} and not on all the other trace samples not used for the analysis attack.
Finally, the profiled correlation attack is computed using the combined values, \emph{i.e.}, the merged traces, exactly as if it were a mono-channel attack.

\begin{figure*}[tb]
\centerline{\includegraphics[width=0.75\textwidth]{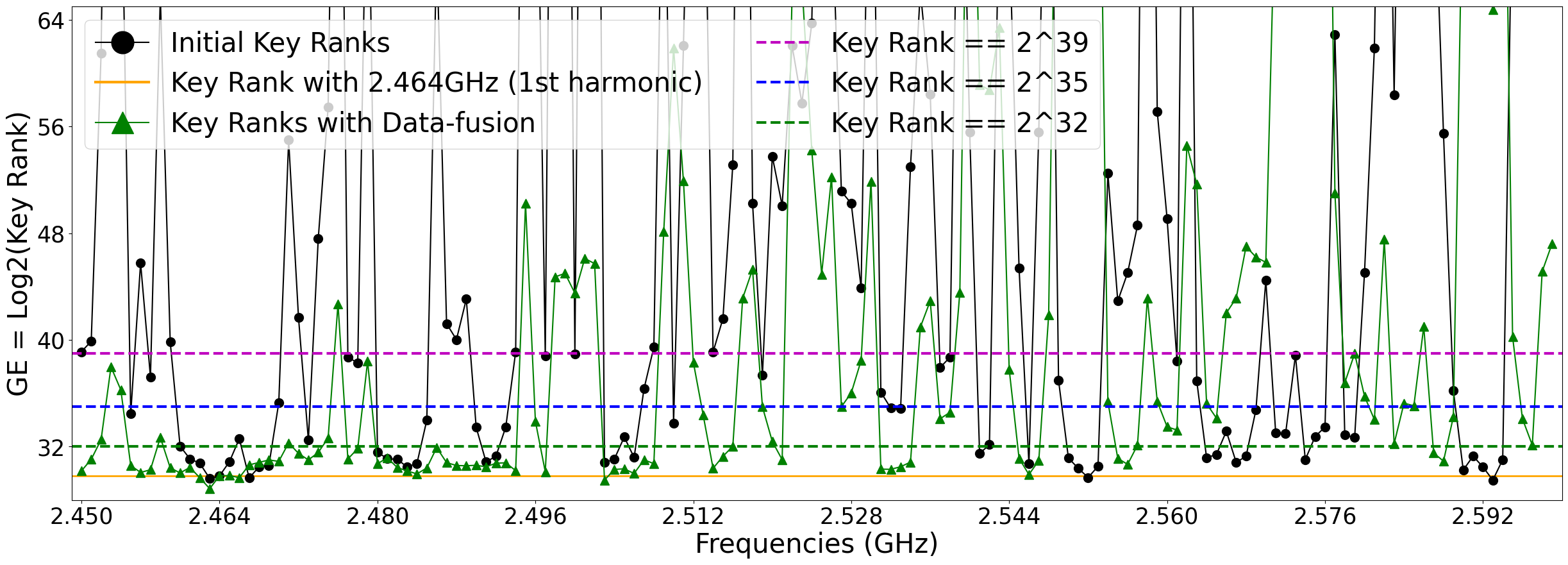}}
\caption{\textbf{Data fusion:} 
Combinations of $2.464$~GHz with $150$ frequencies using data fusion. 
The lower the \ac{ge}, the better the result of the direct combination. 
}
\label{fig:c5:compatible_freqs_464}
\end{figure*}

To analyze the effectiveness of data fusion for frequency combination, we combine the first harmonic at $2.464$~GHz with the $150$~other collected frequencies.
We show the results in Fig.~\ref{fig:c5:compatible_freqs_464}.
The black curve with circle markers corresponds to the attack performance at each of the initial $150$~frequencies, and the orange horizontal line equals the \ac{ge} of the attack at $2.464$~GHz.
The green curve with triangle markers corresponds to the combination results.
Dashed lines indicate different \ac{kr} thresholds.
As a result of the combination, we expect the green curve to be below the black one, which means that combining individual frequencies with the original at $2.464$~GHz improves performance. 
At each frequency, the combination is tested $20$~times, and the results in the figure correspond to the average of the $20$~\acp{ge} from these combined attacks.
Each attack (initial and with combination) is made using $750\times10$ traces (traces$\times$time~diversity).

\begin{figure}[tb]
\centering
\subfloat[]{\includegraphics[width=0.45\textwidth]{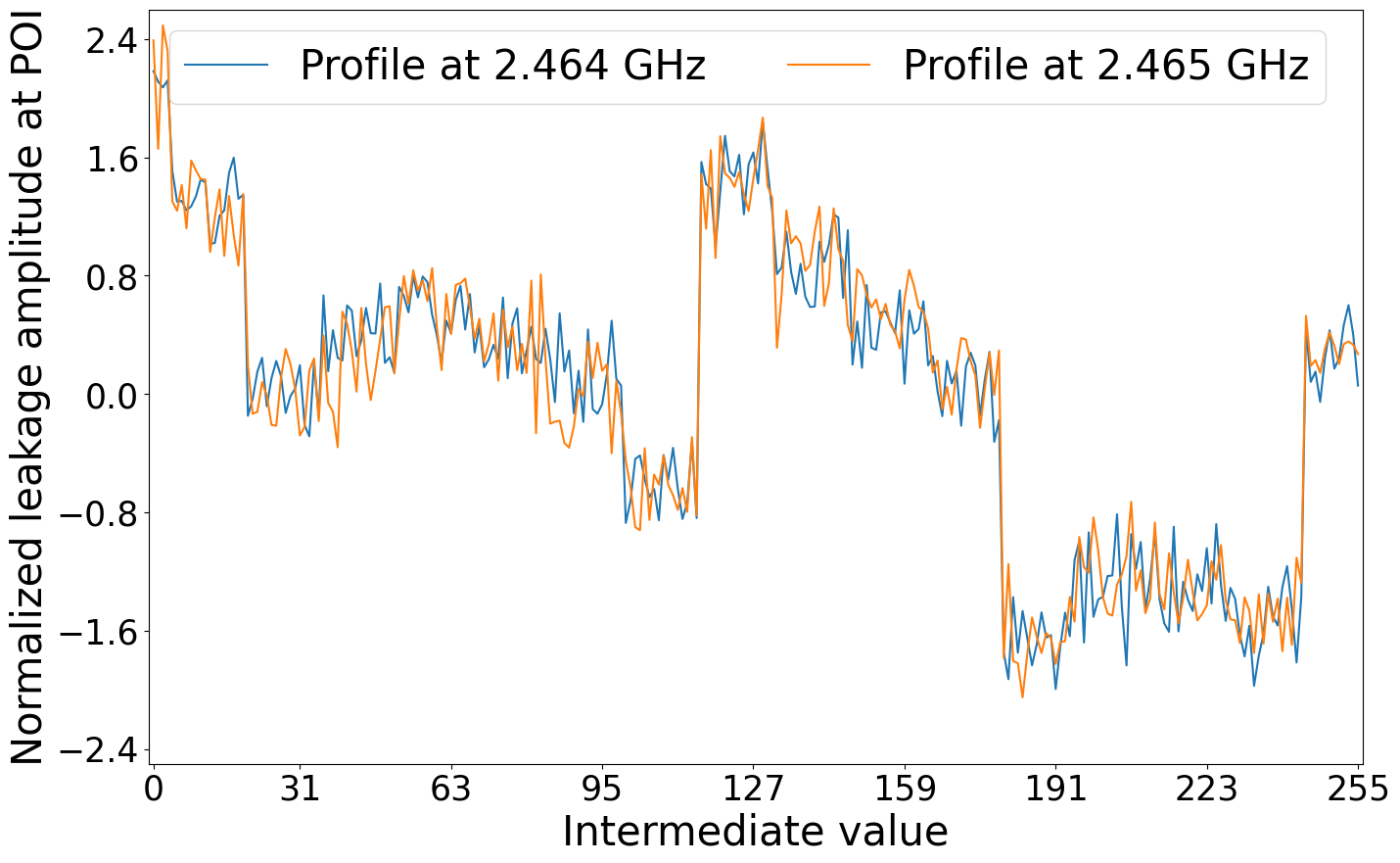}
\label{fig:c5:profile_similarity_2465}}

\subfloat[]{\includegraphics[width=0.45\textwidth]{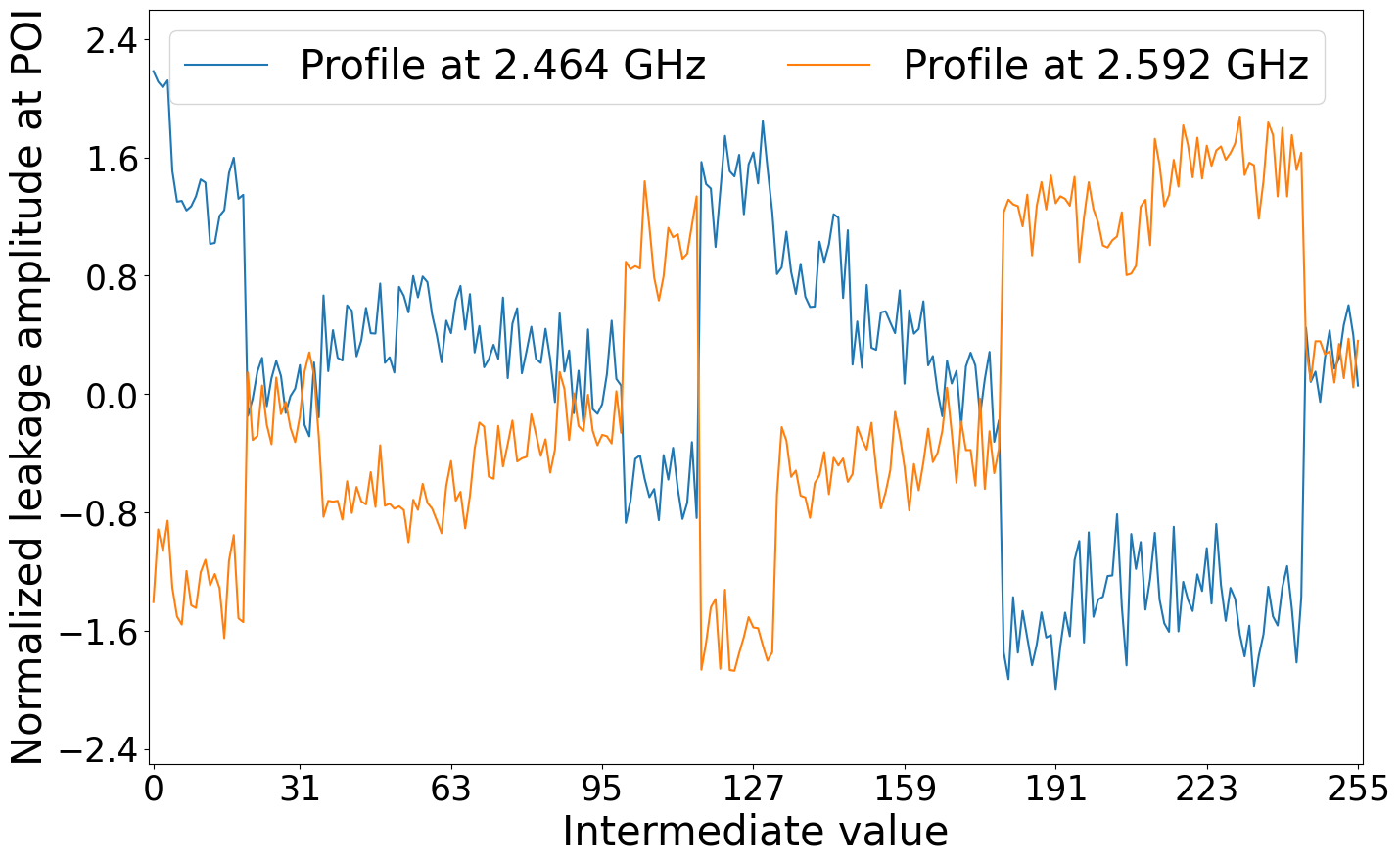}
\label{fig:c5:profile_similarity_2592}}
\caption{\textbf{Profile similarity:} 
Normalized leakage profiles for byte $0$ at (a) two frequencies with similar profiles: $2.464$~GHz and $2.465$~GHz; and (b) two frequencies with inverted leakage profiles: $2.464$~GHz and $2.592$~GHz. }
\label{fig:c5:profile_similarity}
\end{figure}

\textbf{Fig.~\ref{fig:c5:compatible_freqs_464} shows performance improvements for many of the tested frequencies, however, we observe the combination is not efficient all along the spectrum}.
For example, the combination of $2.464$~GHz is working well with $2.465$~GHz and further frequencies like $2.596$~GHz but not with $2.592$~GHz.
To try to explain this behavior, we show the \emph{leakage profiles}, \emph{i.e.}, the profiled leakage amplitude according to the \ac{iv}, across the 256 possible \acp{iv} in Fig.~\ref{fig:c5:profile_similarity}.
When looking at the leakage profile at $2.464$~GHz, we see it is similar to the profile at $2.465$~GHz, as shown in Fig.~\ref{fig:c5:profile_similarity_2465}, which is a frequency where the combination is effective.
On the contrary, the profile is inverted with respect to the one at $2.592$~GHz, as shown in Fig.~\ref{fig:c5:profile_similarity_2592}. 
This problem can be corrected by inverting the values from $2.592$~GHz, which renders the combination effective. 
However, this adds the constraint of checking if two frequencies can be directly combined according to the similarity of their leakage profiles.

The \emph{leakage distortion} observed and analyzed by Camurati et al.~\cite{camurati2020understanding} could explain why the profile differs from one frequency to another. 
Indeed, this distortion, which is expected to happen through the path between leakage generation from the digital part to the \ac{rf} module, can be different from one frequency to another. 
From our observations, this distortion over the spectrum is stable over time, \emph{i.e.}, the same result is obtained when repeating the combination between two identical frequencies multiple times.

\subsubsection{Decision fusion}
\label{sec:c5:decision_fusion}
With the decision fusion method, the combination consists of performing independent mono-channel attacks, one attack for each collected channel, and then combining the attack scores.
The attack chain is exactly the same as in a mono-channel attack, with the addition of an aggregation function used to combine the decisions from the individual attacks.

In a first step, we propose to compare $3$~different aggregation functions used in previous works to combine scores: 
\begin{itemize}
    \item \texttt{average(scores $f1$,  scores $f2$)}~\cite{souissi2012towards}
    \item \texttt{maximum(scores $f1$,  scores $f2$)}~\cite{souissi2012towards}
    \item \texttt{product(scores $f1$,  scores $f2$)}~\cite{camurati2020understanding}
\end{itemize}
To this end, a similar experiment as the one evaluating data fusion over the $150$~frequencies is done with decision fusion. 
To distinguish which of the $3$~aggregation functions is the most efficient, we go into a more challenging situation by selecting a weaker frequency than $2.464$~GHz.
When using $2.464$~GHz, the combinations always return very low \acp{ge}, mainly because $2.464$~GHz initially has a low \ac{ge}, making the contribution from the combination more difficult to observe.

For this experiment, we select the frequency $2.521$~GHz because it has a \ac{ge} of $50$.
A brute-force attack is considered feasible in a reasonable time only if the \ac{ge} is less than or equal to $39$. 
This \ac{ge} threshold value is based on the capacity of our experimental computer to brute-force the key and could vary with the time of brute force considered reasonable by the attacker.
With \ac{ge}=$39$, our computer runs near to $24$~hours to retrieve the key. This time can be decreased with a more powerful computer.
Although a \ac{ge} of $50$ is close to a reasonable \ac{ge}, it is too high for a successful attack since the experimental computer would need at least multiple weeks (even months) to brute-force the key.
Therefore, it is an interesting case study to analyze which aggregation functions make the attack feasible by combining the frequency $2.521$~GHz with other frequencies. 

\begin{figure*}[!htb]

\centering
\subfloat[Aggregation function: \texttt{AVG(scores)}]{\includegraphics[width=0.7\textwidth]{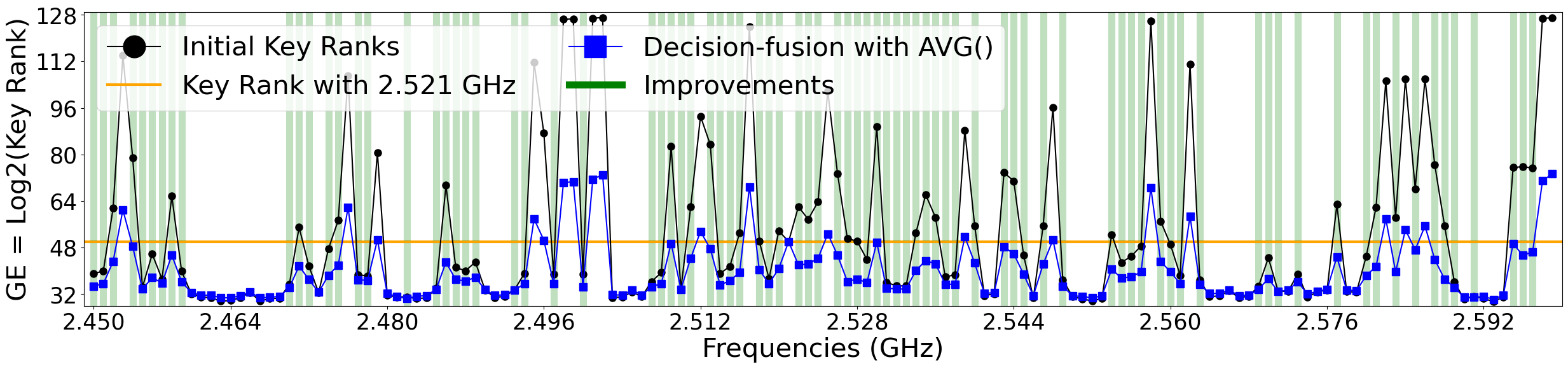}
\label{fig:c5:aggregation_func_AVG}}

\centering
\subfloat[Aggregation function: \texttt{MAX(scores)}]{\includegraphics[width=0.7\textwidth]{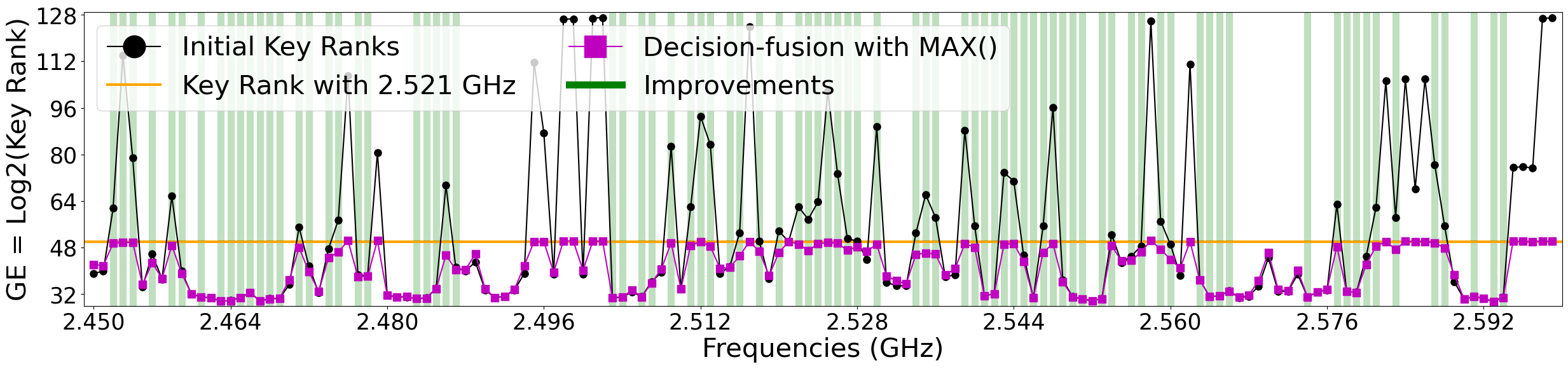} 
\label{fig:c5:aggregation_func_MAX}}

\subfloat[Aggregation function: \texttt{PROD(scores)}]{\includegraphics[width=0.7\textwidth]{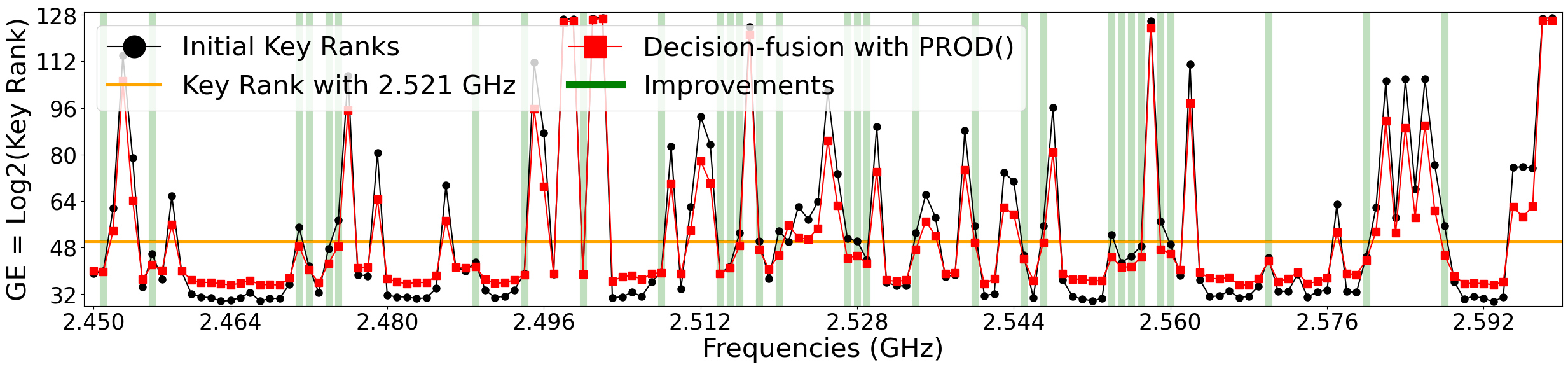}
\label{fig:c5:aggregation_func_PROD}}

\caption{\textbf{Aggregation functions for decision fusion:} Combinations of $2.521$~GHz (orange horizontal line) with the $150$~frequencies (black circles). 
The frequencies where the combination improves attack performance are highlighted in green.
The results for the $3$ aggregation functions are shown in 
(\ref{fig:c5:aggregation_func_AVG}) for the average, 
(\ref{fig:c5:aggregation_func_MAX}) for the maximum and 
(\ref{fig:c5:aggregation_func_PROD}) for the product of scores.}
\label{fig:c5:aggregation_functions}
\end{figure*}

The frequency $2.521$~GHz is combined with the remaining $150$~frequencies using the $3$~tested functions, and combination results are shown in Fig.~\ref{fig:c5:aggregation_functions}.
Every combination result corresponds to the average of $20$~attacks using $750\times10$~traces each.
Table~\ref{tab:c5:aggregation_functions} summarizes the improvement for the $3$ functions.
The improvement equals the lowest \ac{ge} of the two combined frequencies minus the \ac{ge} of the combination. 
The combination improves attack performance when this subtraction is positive, \emph{i.e.}, when the combination reduces \ac{ge} compared to the lowest \ac{ge} of the initial attacks.
To compare the $3$~aggregation functions, the first row of Table~\ref{tab:c5:aggregation_functions} indicates the number of frequencies where the combination brings improvement. 
The second row corresponds to the sum of the improvement over all frequencies where it exists. 

When looking at Fig.~\ref{fig:c5:aggregation_functions}, one could intuitively argue that the \texttt{max()} function performs better as the \ac{ge} never goes higher than $50$.
However, it mostly follows the results of the best frequency being combined and does not bring particular improvement.
As highlighted by the second row of the table~\ref{tab:c5:aggregation_functions}, the \texttt{avg()} function is the one that returns the best improvement over the $150$~combinations.
In particular, it shows a better \ac{ge} reduction in challenging cases where the two combined frequencies have high \ac{ge}.

\newcolumntype{R}[1]{>{\raggedleft\arraybackslash }b{#1}}
\newcolumntype{L}[1]{>{\raggedright\arraybackslash }b{#1}}
\newcolumntype{C}[1]{>{\centering\arraybackslash }b{#1}}
\begin{table}[!tb]
\caption{Comparison of aggregation functions for decision fusion.
}
\centering
\begin{threeparttable}
\begin{tabular}{|C{3.5cm}||C{1cm}|C{1cm}|C{1cm}|}
\hline  Aggregation function     &  \texttt{AVG()}  &   \texttt{MAX()}   &   \texttt{PROD()}   \\
\hline Number of improvements\tnote{a}    &    84   &     83    &     31     \\  
\hline Sum of GE reduction\tnote{b}       &   407   &     99    &     67     \\ 
\hline
\end{tabular}
\label{tab:c5:aggregation_functions}
\begin{tablenotes}
  \item[a] Number of improvements: Corresponds to the number of frequencies where the combination improves performance.
  \item[b] Sum of \ac{ge} reduction: The sum, over the $150$~frequencies, of the difference between the lowest \ac{ge} of combined frequencies and the \ac{ge} of the combination.
\end{tablenotes}
\end{threeparttable}
\end{table}

\begin{figure*}[tb]
\centerline{\includegraphics[width=0.75\textwidth]{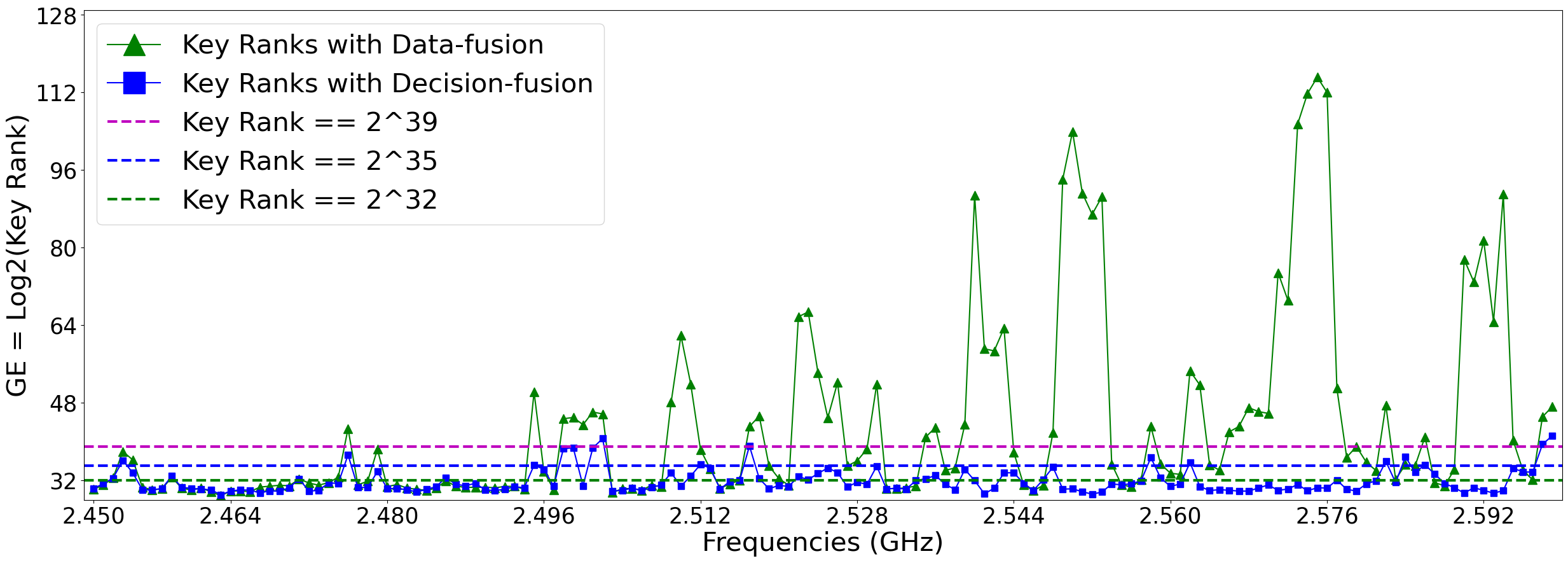}}
\caption{\textbf{Data vs Decision fusion:} 
Combinations of $2.464$~GHz with the $150$~collected frequencies using both data fusion and decision fusion methods. 
The lower the \ac{ge}, the better the result of the direct combination. 
While data fusion does not work with all frequencies, decision fusion returns a low \ac{ge} (below $39$) almost all along the spectrum.
}
\label{fig:c5:data_vs_decision_fusion}
\end{figure*}

In the second step, we compare decision fusion with data fusion by analyzing the results shown in Fig.~\ref{fig:c5:data_vs_decision_fusion}. 
This figure shows the result of a similar experiment as in~\ref{sec:c5:data_fusion}: the combination between the first harmonic ($2.464$~GHz) with the $150$~initial frequencies, but using both data fusion (green line with triangle markers) and decision fusion (with the \texttt{avg()} function) (blue line with square markers).
It highlights the advantage of decision fusion over data fusion in combining multiple frequencies in the context of a screaming-channel attack: while data fusion is inefficient at some frequencies without any adaptation between frequencies, decision fusion fits all along the spectrum.
Additionally, it can be observed that when data fusion is effective, it does not bring better improvement compared to decision fusion. 
Therefore, data fusion and decision fusion have similar performance when both are effective, with the advantage of decision fusion being independent of profile similarity between the combined frequencies, making it more interesting in practice.

In the remainder of this paper, to simplify the analysis, only the results from the most effective combination method are presented: the decision fusion with the \texttt{avg()} function.
However, we notice that it is not essential for an attacker to know which combination method is the most effective in a given case.
It is possible to compute multiple combination methods in parallel and wait until one \ac{ge} from individual attacks or combination methods becomes low enough for a reasonable brute-force attack.

\section{Improving the attack performance with frequency combination}   
\label{sec:att_improvement}

In this section, we highlight the interest in frequency diversity and quantify the improvement that frequency combination can bring in attack performance. 
We perform two studies:
\begin{enumerate}
    \item Combinations of frequencies that are individually too weak for successful attacks. We analyze how increasing the frequency diversity order can make attacks feasible.
    \item An analysis of the combined attack performance according to the initial performance of the combined frequencies. This analysis is done in two cases: combinations of frequencies having equivalent and non-equivalent performance.
\end{enumerate}

\subsection{Making the attack feasible with weak frequencies}
\label{sec:weak_freqs}
In the first study, we consider a \emph{scenario where the attacker can collect a limited number of traces}.
This number of traces is too low to build a successful attack for the best frequency among the $150$, \emph{i.e.}, the \ac{ge} from the best frequency is above $39$.
We investigate the improvement of the attack in these conditions according to the number of combined frequencies.
More specifically, the question is: \textbf{can frequency diversity decrease the \ac{ge} enough to make the attack feasible?}

The number of traces per attack is limited to $20\times10$ (traces$\times$time~diversity).
Under these conditions, the best of the $150$~frequencies has a \ac{ge} close to $50$. 
As explained in Section~\ref{sec:c5:decision_fusion}, with a \ac{ge} of $50$, one can consider the brute-force attack to be unfeasible in a reasonable time; at least multiple weeks would be necessary to retrieve the key. 
With only $20\times10$~traces per attack, there are enough traces in the trace sets to perform $100$~attacks at each frequency.
Therefore, each attack result in this experiment corresponds to the average of $100$~\acp{ge}, guaranteeing strong statistical stability. 

We perform the experiment presented in Algorithm~\ref{alg:c5:alg_comb_order}.
The $150$~frequencies are sorted from best to worst, \emph{i.e.}, from $f_{0}$ having the lowest \ac{ge} to $f_{149}$ having the highest \ac{ge}.
The \ac{ge} of $f_{0}$ equals $49.9$. 
The best frequency $f_{0}$ (included in $Freqs_{combined}$) is combined with the $150$~initial frequencies (lines $5$~--~$8$). 
The best of the $1^{st}$ order combinations get a \ac{ge} equal to $41.8$.
This \ac{ge} value is still very high as it would require at least one week to brute-force the key. 
Nevertheless, it is getting closer to a reasonable brute-force attack.

The best $1^{st}$ order combination corresponds to the one between $f_{0}$ and the second best frequency $f_{1}$.
Therefore, $f_{1}$ is added to the list $Freqs_{combined}$  (line $9$).
To increase the frequency diversity order, another iteration is computed (lines $3$~--~$11$)  
to combine the frequencies included in $Freqs_{combined}$: \emph{e.g.},  $f_{0}$ and $f_{1}$ after the first iteration, with the $150$~initial frequencies. 
Frequency diversity is increased up to the $4^{th}$ order, and the best \ac{ge}, initially equal to $49.9$, goes as low as $36.4$, making a brute-force attack possible in a few hours.

\begin{algorithm}[tb]
\caption{Finding the best combination for each frequency diversity order}\label{alg:c5:alg_comb_order}
\begin{algorithmic}[1]
    \REQUIRE Scores of the $150$~frequencies: $f_{i}$ with $i \in [0; 149]$ \COMMENT{frequencies sorted from best to worst}
    \REQUIRE $Limit\_FreqDiv_{order}$  \COMMENT{Required frequency diversity order}
    \ENSURE $Lowest\_GE$
    \STATE $Freqs_{combined}$ = [$f_{0}$]
    \STATE $FreqDiv_{order}  \leftarrow 1  $    
    \REPEAT 
        \STATE $i  \leftarrow 0  $
        \REPEAT 
            \STATE $GEs\_Combination[\textbf{\textit{i}}] = \newline Decision\_Fusion(Freqs_{combined} \text{ \& } f_{\textbf{\textit{i}}})$
            \STATE $i \leftarrow i + 1 $
        \UNTIL {$i >= 150$}
        \STATE $Freqs_{combined}.append( ArgMin(GEs\_Combination))$ \COMMENT{the frequency returning the best combination is added to the list of frequencies to combine}
        \STATE $FreqDiv_{order} \leftarrow FreqDiv_{order} + 1 $
    \UNTIL {$FreqDiv_{order} > Limit\_FreqDiv_{order}$}
    \STATE $Lowest\_GE \leftarrow Min(GEs\_Combination)$

\end{algorithmic}
\end{algorithm}

\begin{figure}[tb]
\centerline{\includegraphics[width=0.5\textwidth]{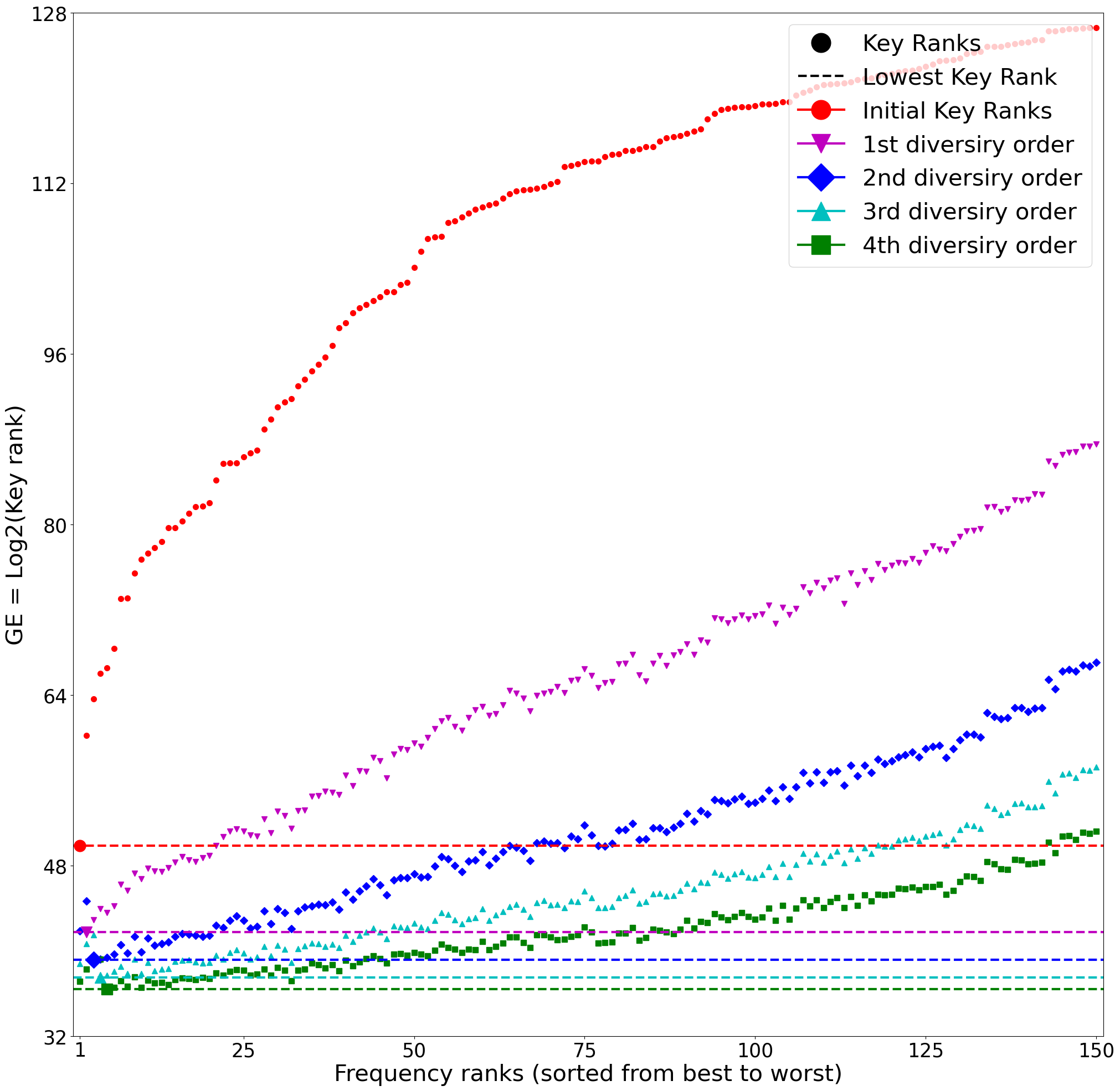}}
\caption{\textbf{\acf{ge} according to frequency diversity order:} 
(Red) The $150$~frequencies are sorted according to their \ac{ge} (red points) from the best to the worst. 
The number of traces is low enough for the best frequency to have a \ac{ge} close to $50$. With $20\times10$~traces, the best \ac{ge} equals $49.9$ (red horizontal dashed line).
(Purple) First-order diversity: the combination of the best frequency with all others (purple points). The new lower \ac{ge} equals $41.8$ (purple horizontal dashed line).
(Blue) Second-order diversity: best combination from first-order combined with all initial frequencies, best \ac{ge} = $39.2$. 
(Light blue) Third-order diversity, best \ac{ge} = $37.5$.
(Green) Fourth-order diversity, best \ac{ge} = $36.4$.
}
\label{fig:c5:combination_order}
\end{figure}

These experimental results are shown in Fig.~\ref{fig:c5:combination_order} summarized in Table~\ref{tab:c5:combination_order}, showing how \textbf{frequency diversity can make attacks feasible when the number of \acp{cp} computed by the victim is too low for a successful mono-channel attack}. 
In essence, frequency diversity allows for increasing the number of traces by collecting traces at different frequencies.
As we could expect, each time the diversity order is increased, the best combination is obtained with the next best frequency that is not included in $Freqs_{combined}$ yet.
At the end of the $4^{th}$ iteration, $Freqs_{combined} = [f_{0}:f_{1}: f_{2}:f_{3}:f_{4}]$, with frequencies $f_{i}$ by order of addition in $Freqs_{combined}$.
One observation is that the more we increase the diversity order, the lower the new \ac{ge} reduction is.
A potential explanation is that most of the information carried by the new frequencies has already been brought by previous frequencies.
We thus observe a law of diminishing return, while the increase in the setup complexity is linear.
Indeed, collecting an additional frequency simultaneously increases the setup complexity since an additional \ac{sdr} must be used.

\begin{table}
\caption{\textbf{\ac{ge} according to frequency diversity order.} 
}
\centering
\begin{threeparttable}
\resizebox{\columnwidth}{!}{ 
\begin{tabular}{@{}cccccc@{}}
\hline  Diversity order  &  $0$  &   $1^{st}$   &   $2^{nd}$ & $3^{rd}$ & $4^{th}$  \\
\hline  $Freqs_{combined}$  
    & $\mathit{f_{0}}$   
    & $\mathit{f_{0}\mathord{:}f_{1}}$
    & $\mathit{f_{0}\mathord{:}f_{1}\mathord{:}f_{2}}$
    & $\mathit{f_{0}\mathord{:}f_{1}\mathord{:}f_{2}\mathord{:}f_{3}}$
    & $\mathit{f_{0}\mathord{:}f_{1}\mathord{:}f_{2}\mathord{:}f_{3}\mathord{:}f_{4}}$ \\ 
\hline  GE &     $49.9$   &   $41.8$      &  $39.2$     &   $37.5$  &  $36.4$ \\   
\hline
\end{tabular}
\label{tab:c5:combination_order}
}
\begin{tablenotes}
  \item[] $f_{i}$: $i^{th}$ frequency among the $150$~frequencies, sorted according to their\\individual performance.
\end{tablenotes}
\end{threeparttable}
\end{table}

\subsection{Combination efficiency according to the performance of the combined frequencies}
\label{sec:combination_efficiency}
In this second study, we evaluate the improvement provided by frequency diversity according to the performance of the initial frequencies.
We analyze under which conditions it is interesting to collect and combine frequencies together.
To simplify the analysis, we focus on the combinations of only two frequencies.
Nevertheless, our results can theoretically be generalized to any number of combined frequencies.
Two cases are identified:
\begin{enumerate}
    \item Combination of two equivalent frequencies, \emph{i.e.}, frequencies with equivalent \acp{ge}.
    \item Combination of two non-equivalent frequencies, \emph{i.e.}, frequencies with different \acp{ge}. We try to answer the question: \textbf{How does the weakest contribute to increasing or decreasing the performance of the strongest?}
\end{enumerate}
Table~\ref{tab:c5:eq_freqs} gives results for the first case, and Table~\ref{tab:c5:ineq_freqs} for the second.
Each result of initial and combined attacks in these tables corresponds to an average of $20$~attacks using $750\times10$ traces each.
To evaluate the combination improvement, both tables give, for a fixed number of traces ($750\times10$), the \ac{ge} of each combined frequency and the \ac{ge} of the combination.

The first observation is that the higher the \acp{ge} of combined frequencies, the more improvement the combination brings.
Starting from the first row, in Table~\ref{tab:c5:eq_freqs}, \acp{ge} from initial frequencies increase by $5$ between most rows, except for the last two rows, where it increases by $10$.
Regarding the \ac{ge} of the combination, it increases at a slower pace: between $2$ and $3$, and it increases by $8$ between the last two rows.
Therefore, when the \acp{ge} from combined frequencies increase, the difference between their \ac{ge} and the \ac{ge} of the combination increases, too.

A second observation when looking at both Table~\ref{tab:c5:eq_freqs} and~\ref{tab:c5:ineq_freqs} is that the combination of equivalent frequencies, having each a $\ac{ge} = ge$, returns a similar result as the combination of non-equivalent frequencies with $Avg(\acp{ge}) = ge$.
For example, combining two non-equivalent frequencies with \acp{ge} of $40$ and $50$ is similar to combining two equivalent frequencies having \acp{ge} of $45$.

Additionally, the tables indicate the number of traces to reach given \acp{ge} of $39$ and $35$.
We observe that the combination is also efficient when the initial frequencies are good enough to reach these \acp{ge} with less than $750\times10$ traces (rows $1\&2$ in Table~\ref{tab:c5:eq_freqs}). 
Frequency diversity reduces the number of traces needed to obtain these \acp{ge}.

The key takeaways of these observations are: (1) \textbf{how frequencies otherwise considered useless for carrying too weak of a leakage component can be combined together to build a successful attack}; and (2) \textbf{how the combination is still efficient with strong frequencies, improving the required number of traces to reach a given \ac{ge}.}

\begin{table*}
\caption{\textbf{Combinations of two equivalent frequencies:} 
Combination of $2$ frequencies having a similar performance, \emph{i.e.}, a similar \ac{ge} when using the same number of traces ($750\times10$ traces here).
}
\centering
\begin{threeparttable}
\begin{tabular}{|C{1.3cm}|C{1.3cm}||C{1.3cm}|C{1.3cm}||C{1.3cm}|C{1.3cm}|C{1.3cm}||C{1.3cm}|C{1.3cm}|}
\hline \multicolumn{2}{|c||}{Frequency 1} & \multicolumn{2}{c||}{Frequency 2} & \multicolumn{3}{c||}{Combination} & \multicolumn{2}{c|}{Best frequency} \\
\hline \multirow{2}{1.3cm}{\centering Freq\\ (GHz)}   & \multirow{2}{1.3cm}{\centering GE}    & \multirow{2}{1.3cm}{\centering Freq\\ (GHz)}      & \multirow{2}{1.3cm}{\centering GE}   & \multirow{2}{1.3cm}{\centering GE}    & \multicolumn{2}{c||}{$min_{traces}$ for GE}   &  \multicolumn{2}{c|}{$min_{traces}$ for GE}\\
\cline{6-9}  &      &        &         &       & $<$ 39  & $<$ 35  & $<$ 39  & $<$ 35  \\
\hline 2.552 & 29.7 &  2.593 &  29.5   &  29.3 & 36      & 52      & 50      &   76  \\  
\hline 2.470 & 35.3 &  2.532 &  34.9   &  32.2 & 185     & 240     & 405     &  719  \\ 
\hline 2.488 & 40.0 &  2.508 &  39.5   &  34.5 & 400     & 645     & $>$750  & $>$750 \\
\hline 2.456 & 45.8 &  2.545 &  45.4   &  37.5 & 645     & $>$750  & $>$750  & $>$750 \\ 
\hline 2.521 & 50.1 &  2.560 &  49.1   &  39.7 & $>$750  & $>$750  & $>$750  & $>$750 \\
\hline 2.452 & 61.5 &  2.511 &  62.1   &  47.4 & $>$750  & $>$750  & $>$750  & $>$750 \\ 
\hline
\end{tabular}
\label{tab:c5:eq_freqs}
\begin{tablenotes}
  \item[] GE: Guessing entropy. 
  \item[] $min_{traces} < 39\ \&\ 35$: minimum number of traces required for the \ac{ge} to be lower than $39\ \&\ 35$ respectively.
\end{tablenotes}
\end{threeparttable}
\end{table*}

\begin{table*}
\caption{\textbf{Combinations of two non-equivalent frequencies:}
Combination of $2$ frequencies not having a similar performance, \emph{i.e.}, different \acp{ge} when using the same number of traces ($750\times10$ traces here).
}
\centering
\begin{threeparttable}
\begin{tabular}{|C{1.3cm}|C{1.3cm}||C{1.3cm}|C{1.3cm}||C{1.3cm}|C{1.3cm}|C{1.3cm}||C{1.3cm}|C{1.3cm}|}
\hline \multicolumn{2}{|c||}{Frequency 1} & \multicolumn{2}{c||}{Frequency 2} & \multicolumn{3}{c||}{Combination} & \multicolumn{2}{c|}{Best frequency} \\
\hline \multirow{2}{1.3cm}{\centering Freq\\ (GHz)}   & \multirow{2}{1.3cm}{\centering GE}    & \multirow{2}{1.3cm}{\centering Freq\\ (GHz)}      & \multirow{2}{1.3cm}{\centering GE}   & \multirow{2}{1.3cm}{\centering GE}    & \multicolumn{2}{c||}{$min_{traces}$ for GE}   &  \multicolumn{2}{c|}{$min_{traces}$ for GE}\\
\cline{6-9}  &      &        &         &       & $<$ 39  & $<$ 35  & $<$ 39  & $<$ 35  \\
\hline 2.470 & 35.3 &  2.488 &  40.0   &  33.6 & 252     &  500    & 431     & $>$750  \\ 
\hline 2.470 & 35.3 &  2.456 &  45.8   &  33.3 & 283     &  531    & 431     & $>$750  \\ 
\hline 2.470 & 35.3 &  2.521 &  50.1   &  34.2 & 310     &  639    & 431     & $>$750  \\
\hline 2.488 & 40.0 &  2.456 &  45.8   &  35.2 & 425     & $>$750  & $>$750  & $>$750  \\ 
\hline 2.488 & 40.0 &  2.521 &  50.1   &  36.5 & 499     & $>$750  & $>$750  & $>$750  \\
\hline 2.488 & 40.0 &  2.452 &  61.5   &  39.6 & $>$750  & $>$750  & $>$750  & $>$750  \\ 
\hline
\end{tabular}
\label{tab:c5:ineq_freqs}
\begin{tablenotes}
  \item[] GE: Guessing entropy. 
  \item[] $min_{traces} < 39\ \&\ 35$: minimum number of traces required for the \ac{ge} to be lower than $39\ \&\ 35$ respectively.
\end{tablenotes}
\end{threeparttable}
\end{table*}

\section{Attacking in more realistic conditions}
\label{sec:realistic_attacks}
The limitation of our studies in the previous section is that traces from the different frequencies are collected by cable and at different times.
In this Section \textbf{we demonstrate the benefits of frequency diversity for multi-screaming channel attacks and verify our observations in more realistic conditions}.
We simultaneously collect leakage at two different frequencies using $2$~\acp{sdr} and \textbf{build successful attacks at challenging distances of $15$ and, for the first time reported in the literature, $30$ meters}.
Profiles are built during a profiling phase on one instance of the nRF52832, collecting leakage by cable.
During the attack phase, a directional antenna collects leakage at a distance of multiple meters from the victim, which is another instance of the same device.

\subsection{Attack results at 15 meters}
\label{sec:attack_15m}
To compare our results with previous works on screaming-channel attacks at $15$~meters~\cite{camurati2020understanding, wang2020far, wang2021advanced}, we reproduce the attack under similar conditions as Camurati et al.~\cite{camurati2020understanding}: $10000\times500$ profiling traces and $5000\times500$ attacking traces are collected at each selected frequency.
However, a proper comparison of attack performance is difficult 
as the conditions differ: environment, \ac{rf} pollution, etc.
As shown in Fig~\ref{fig:c5:attack_results_a}, when using the second harmonic ($2.528$~GHz), the results differ: 
The \ac{ge} from Camurati et al. decreases slowly but continuously.
On the contrary, our \ac{ge} decreases fast at first but then stabilizes and does not strictly decrease, as we can see in Fig.~\ref{fig:c5:attack_results_a} between traces $1000$ and $2000$ and also between traces $3000$ and $4000$.
This is probably because this frequency was partially polluted in our environment. Thus, some traces decrease the attack performance, likely because they mostly collect noise. 
It is also difficult to make a proper comparison with works from Wang et al. ~\cite{wang2020far, wang2021advanced} since they evaluate their attack on the recovery of only one key byte.
Therefore, we evaluate the combination by comparing its performance with the individual performance of the combined frequencies.

To consider a realistic use case, and since the second harmonic leakage is very strong when not polluted, we consider a situation where this frequency would be too polluted so the attacker would have to use other non-harmonic frequencies.
We select the two non-harmonic frequencies that performed the best at $7$~meters in the works exploring leakage at non-harmonics~\cite{guillaume2023attacking}: $2.484$~GHz and $2.593$~GHz, and compare the result of the combination with the attack performance at each individual frequency.
Fig.~\ref{fig:c5:attack_results_b} shows the results of both individual frequencies and the result of their combination according to the number of attack traces.
Each attack result is the average of $5$~attacks using $1000\times500$ traces.
Table in Fig.~\ref{fig:c5:final} summarizes the performance of each frequency and their combination.
To get a \ac{ge} of $32$, the two frequencies need $410\times500$ and $686\times500$ traces.
When combined, $136\times500$ traces are enough to get the same result.
We can see that their combination improves the attack by decreasing the number of necessary traces to reach the lowest \ac{ge}. 
However, it does not decrease the reachable \ac{ge}. 
When both frequencies reach a limit, their combination does not bring any improvement but stabilizes around $32$.
We hypothesize that these non-harmonic frequencies transmit information on fewer bits than the second harmonic. 
Then, even decreasing the noise level by combining the two frequencies cannot allow the recovery of information not carried at these frequencies.

\subsection{Attack results at 30 meters}
\label{sec:attack_30m}
Going beyond previous works, we now attack at a larger distance of $30$~meters, where it is more difficult to find frequencies with observable leakage patterns.
In the environment where this experiment is performed, the second harmonic is not polluted.
We select $2.528$~GHz and $2.552$~GHz by empirically testing frequencies where the traces correspond to the leakage pattern. 
Our observation highlights a potential interest in harmonics compared to non-harmonics, which is that leakage seems to be visible at further distances at these frequencies.
To have a better resolution on the number of traces needed to get a given \ac{ge}, we reduce the time diversity compared to the attack at $15$~meters and go from $500$ to $50$.

\begin{figure*}[tb]
\centering
\subfloat[Initial mono-channel attacks at 15 meters with the second harmonic ($2.528$~GHz)]{\includegraphics[width=0.75\textwidth]{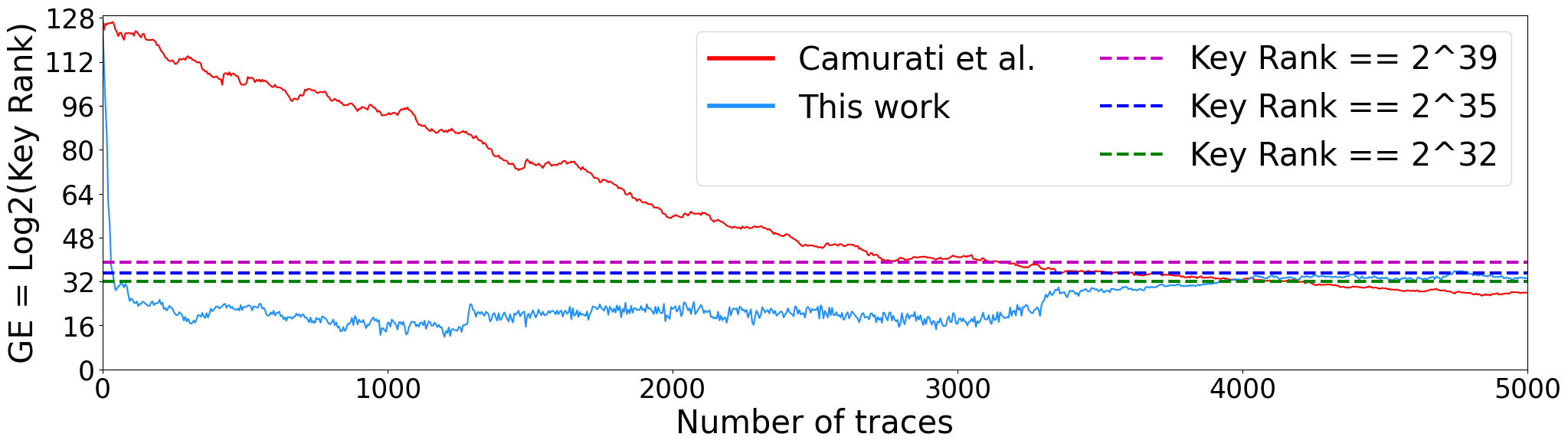}\label{fig:c5:attack_results_a}}

\subfloat[Multi-channel attack at 15 meters]{\includegraphics[width=0.45\textwidth]{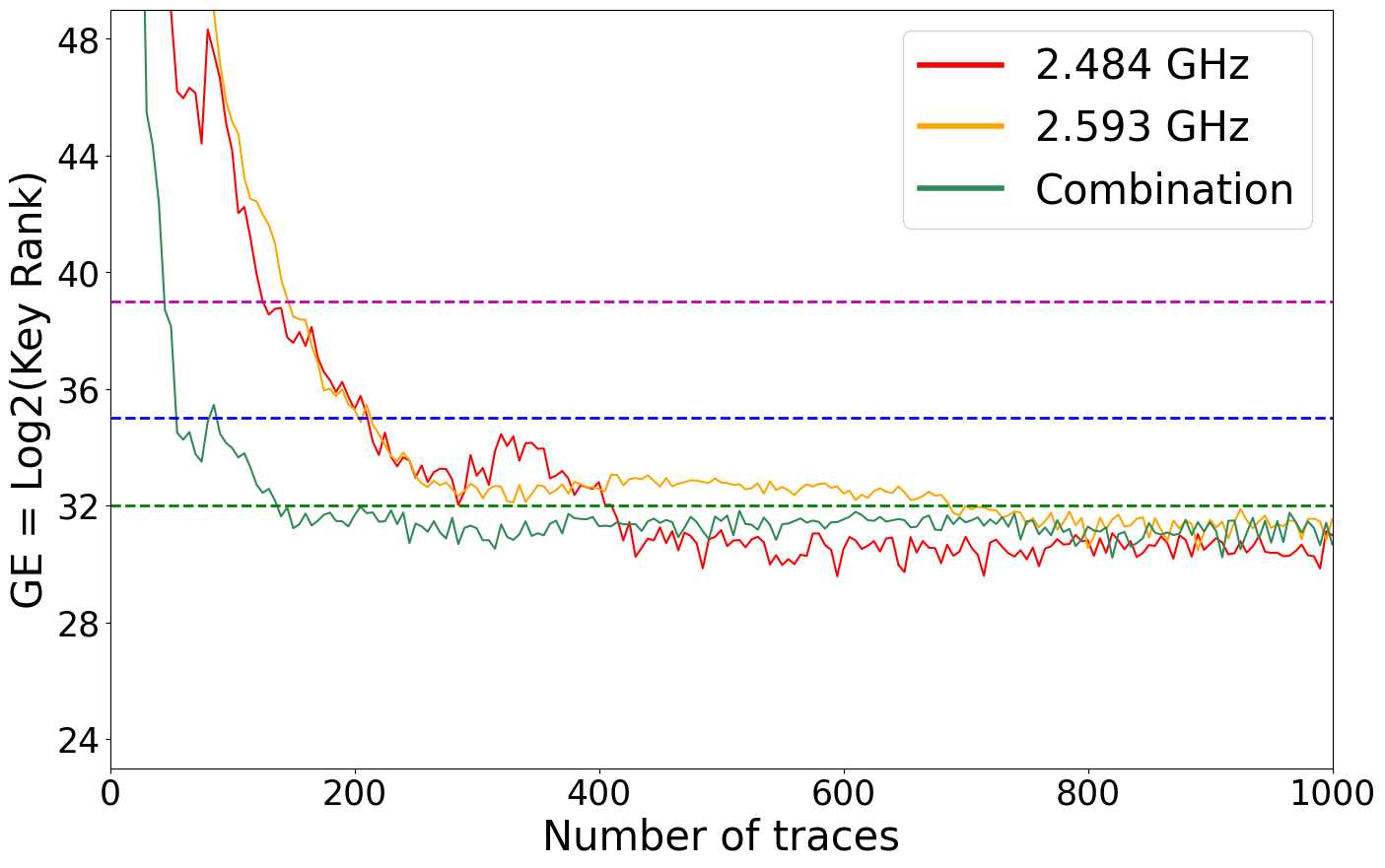}
\label{fig:c5:attack_results_b}}
\hspace{0.6cm}
\subfloat[Multi-channel attack at 30 meters]{\includegraphics[width=0.45\textwidth]{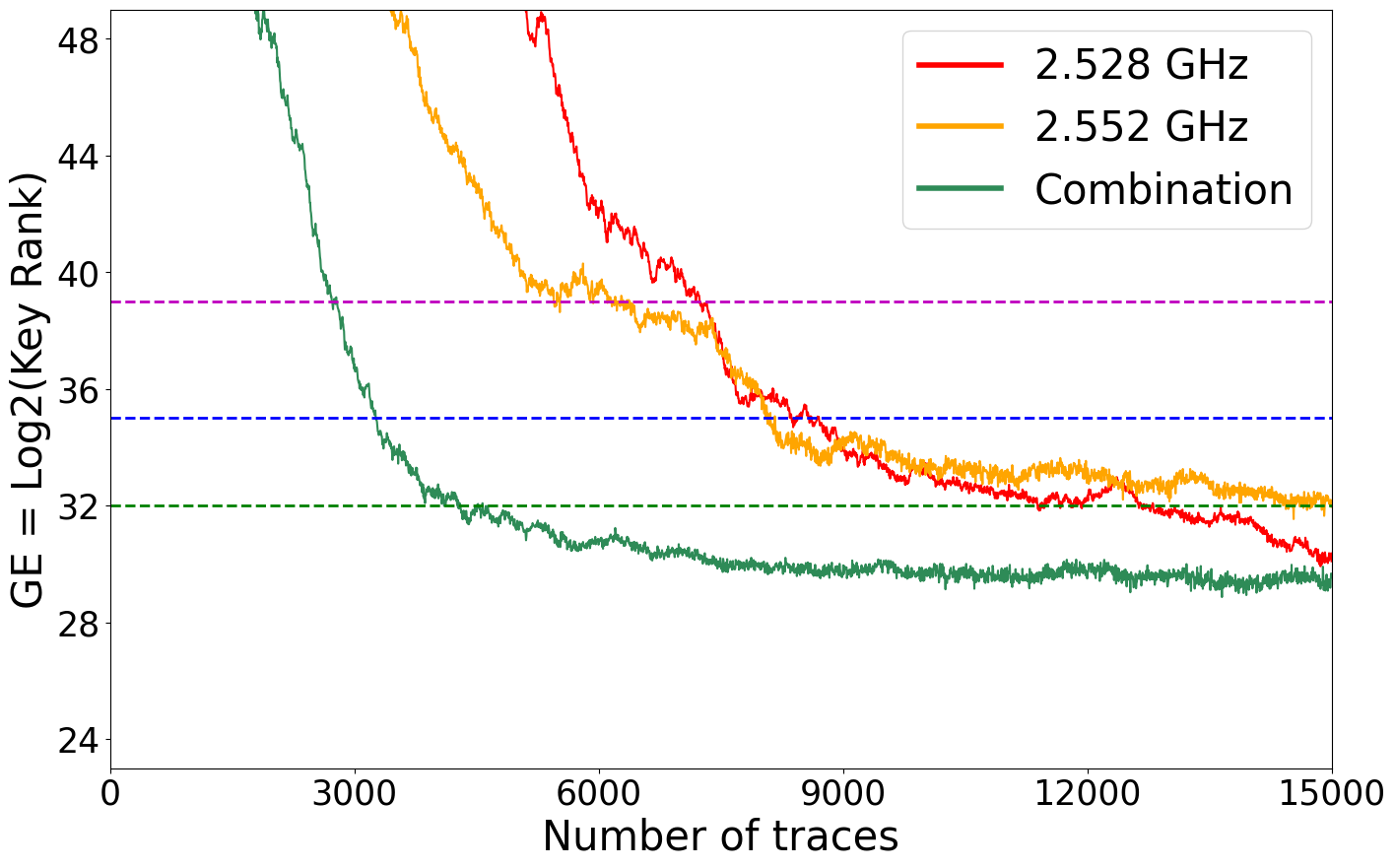}
\label{fig:c5:attack_results_c}}

\vspace{1ex}
\resizebox{1.1\columnwidth}{!}{
\begin{tabular}{|C{1.3cm}|C{2cm}||C{2cm}|C{2cm}|C{2cm}|}
\hline\multirow{2}{1.3cm}{\centering Distance}    & \multirow{2}{2cm}{\centering Frequency}      &  \multicolumn{3}{c|}{$min_{traces}$ for GE} \\
\cline{3-5}                                                      &     & \cellcolor{violet!50}    $<$39         &  \cellcolor{blue!50}      $<$35       & \cellcolor{green!50}  $<$32       \\
\hline       \rowcolor{red!70}    \cellcolor{white}                                                 &     $2.484$~GHz       &  $125\times500$       &  $210\times500$    &  $410\times500$       \\  
\cline{2-5}  \rowcolor{orange!70} \cellcolor{white}                                                 &     $2.593$~GHz       &  $146\times500$      &  $203\times500$    &  $686\times500$       \\  
\cline{2-5}  \rowcolor{green!70}  \cellcolor{white}  \multirow{-3}{1.3cm}{\centering 15 meters}     &  Combination          &  $ 45\times500$      &  $ 54\times500$    &  $136\times500$       \\   
\hline       \rowcolor{red!70}    \cellcolor{white}                                                 &     $2.528$~GHz       &  $7\ 242\times50$    &  $8\ 370\times50$  &  $11\ 390\times50$       \\  
\cline{2-5}  \rowcolor{orange!70} \cellcolor{white}                                                 &     $2.552$~GHz       &  $5\ 452\times50$    &  $8\ 070\times50$  &  $14\ 439\times50$       \\  
\cline{2-5}  \rowcolor{green!70}  \cellcolor{white}  \multirow{-3}{1.3cm}{\centering 30 meters}     &  Combination          &  $2\ 727\times50$    &  $3\ 264\times50$  &  $ 4\ 105\times50$       \\   
\hline
\end{tabular}
}
\label{tab:c5:attack_results}
\caption{\textbf{Experimental results:} 
(a) Comparison of attacks at $15$ meters using the second harmonic ($2.528$~GHz) in (red) Camurati et al. conditions~\cite{camurati2020understanding} and (blue) this work conditions.
Initial frequencies vs their combination at (b) $15$ meters and (c) $30$ meters.
}
\label{fig:c5:final}
\end{figure*}

Fig.~\ref{fig:c5:attack_results_c} shows the results vs. the number of traces.
Here again, each result corresponds to the average of $5$~attacks in the same conditions.
We can clearly see how the combination divides by $2$ the number of traces needed to get a \ac{ge} lower than $39$ compared to the best of the initial frequencies.

\section{Discussion}
\label{sec:discusion}
In previous studies on screaming-channel attacks, increasing the number of collected traces is the only solution to reduce the \ac{ge} until it becomes low enough for a brute-force attack to be possible in a reasonable time. 
In our experiments for \textbf{multi-screaming channel attacks}, by increasing the frequency diversity order, we demonstrated a successful attack in a situation where the number of \acp{cp} executed by the victim is not enough for any mono-channel attack to succeed.
Therefore, \textbf{increasing the frequency diversity order for screaming-channel attacks by collecting traces at different frequencies can effectively increase the number of collected traces without having the victim execute more \acp{cp}}.
We have shown how even if these traces contain leakage from the same \ac{cp} executions, their combination increases attack performance.
Our results contribute to a \textbf{more realistic threat model} where a lower number of \ac{cp} executions is needed for a successful attack.

We confirm, at a distance of $15$~meters, the results from Guillaume et al~\cite{guillaume2023attacking} that attacking at non-harmonics can perform similarly to attacking at harmonics.
The results also highlight how the combination of non-harmonics, in a scenario where harmonics would be polluted, makes it possible to get closer to the performance obtained with harmonics.

Our study considers a \ac{ge} to be reasonable for a brute-force attack when equal to or less than $39$, making a brute-force attack possible in less than $24$~hours. 
In our experiments, combining two frequencies with \acp{ge} close to $50$ returns a \ac{ge} very close to $39$.
Then, an attack with frequency diversity performs similarly to a mono-channel attack whose reasonable \ac{ge} would be $50$ instead of $39$.
With a more powerful computer capable of brute-forcing a key with higher \ac{ge}, \emph{e.g.}, $48$, according to our results, it would be possible to succeed in an attack with two frequencies having individual GE of $60$.
The results of this paper increase the number of exploitable frequencies.
Since weaker frequencies with higher \acp{ge} can make the attack possible when combined together, the surface of potential frequencies to use is increased again.

However, we did not demonstrate that frequency diversity can lower the reachable \ac{ge}, \emph{i.e.}, the limit to which the \ac{ge} stabilizes to when enough traces are collected.
Our hypothesis, whose study is kept for future work, is that the combined frequencies do not carry information on all key bits.
Frequency diversity increases the \ac{snr} by reducing the noise, but it cannot solve this lack of information.

A limitation of the first studies in this paper is that they are done with traces collected at different times.
However, we assume the results obtained in these conditions are similar if frequencies were collected simultaneously since the frequencies are spaced enough to be in different coherence zones and thus expected to carry independent noise.
We verified this hypothesis in our last experiment with leakage from common \acp{cp} collected simultaneously at two frequencies and whose combination improves attack performance.

\section{Conclusion}
\label{sec:conclusion}
In this paper, we have studied the interest of frequency diversity in the context of screaming-channels attacks and proposed multi-screaming channel attacks.
We have demonstrated the effectiveness of the possible methods to combine frequencies and analyzed the conditions under which this diversity is the most interesting.

Two combination methods are studied: data fusion and decision fusion.
To compare them, we combined the second harmonic with $150$~frequencies along the considered range of the spectrum and showed that decision fusion is superior to data fusion as the combination works throughout the spectrum.
On the contrary, data fusion depends on profile similarity between combined frequencies. 

To study the improvement of multi-screaming channel attacks, we proposed two case studies.
First, we demonstrated the improvement in attack performance in a situation where the number of traces is too low to mount a successful mono-channel attack.
In these conditions, frequency diversity keeps the attack feasible.
Second, we quantify the improvement according to the performance of the combined frequencies by looking into the reduction of both the \ac{ge} and the number of traces needed to get exploitable \acp{ge}.
We demonstrate that combinations of both equivalent and non-equivalent frequencies are of interest.
Finally, we show the interest of frequency diversity in realistic cases with attacks performed at $15$ and, for the first time reported, at $30$~meters, collecting the leakage simultaneously at two frequencies.

In this work, when multiple frequencies are combined, they all have the same weight in the final decision, even when combining two non-equivalent frequencies.
One potential future work is to investigate the addition of weights for each frequency according to their respective performance.
Also, we only focus on the right side of the spectrum with respect to the legitimate signal at $2.4$~GHz.
Another possible line of work would be to study the combination of symmetric frequencies: \emph{i.e.}, $2.4$~GHz + $f$ and $2.4$~GHz - $f$.


\begin{thebibliography}{10}
\providecommand{\url}[1]{#1}
\csname url@samestyle\endcsname
\providecommand{\newblock}{\relax}
\providecommand{\bibinfo}[2]{#2}
\providecommand{\BIBentrySTDinterwordspacing}{\spaceskip=0pt\relax}
\providecommand{\BIBentryALTinterwordstretchfactor}{4}
\providecommand{\BIBentryALTinterwordspacing}{\spaceskip=\fontdimen2\font plus
\BIBentryALTinterwordstretchfactor\fontdimen3\font minus \fontdimen4\font\relax}
\providecommand{\BIBforeignlanguage}[2]{{%
\expandafter\ifx\csname l@#1\endcsname\relax
\typeout{** WARNING: IEEEtran.bst: No hyphenation pattern has been}%
\typeout{** loaded for the language `#1'. Using the pattern for}%
\typeout{** the default language instead.}%
\else
\language=\csname l@#1\endcsname
\fi
#2}}
\providecommand{\BIBdecl}{\relax}
\BIBdecl

\bibitem{choi2020tempest}
J.~Choi, H.-Y. Yang, and D.-H. Cho, ``Tempest comeback: A realistic audio eavesdropping threat on mixed-signal socs,'' in \emph{Proceedings of the 2020 ACM SIGSAC Conference on Computer and Communications Security}, 2020.

\bibitem{standaert2010introduction}
F.-X. Standaert, ``Introduction to side-channel attacks,'' \emph{Secure integrated circuits and systems}, 2010.

\bibitem{mangard2008power}
S.~Mangard, E.~Oswald, and T.~Popp, \emph{Power analysis attacks: Revealing the secrets of smart cards}.\hskip 1em plus 0.5em minus 0.4em\relax Springer Science \& Business Media, 2008, vol.~31.

\bibitem{gandolfi2001electromagnetic}
K.~Gandolfi, C.~Mourtel, and F.~Olivier, ``Electromagnetic analysis: Concrete results,'' in \emph{Cryptographic Hardware and Embedded Systems--CHES}.\hskip 1em plus 0.5em minus 0.4em\relax Springer, 2001.

\bibitem{camurati2018screaming}
G.~Camurati, S.~Poeplau, M.~Muench, T.~Hayes, and A.~Francillon, ``Screaming channels: When electromagnetic side channels meet radio transceivers,'' in \emph{Proceedings of the 2018 ACM SIGSAC Conference on Computer and Communications Security}, 2018.

\bibitem{guillaume2023attacking}
J.~Guillaume, M.~Pelcat, A.~Nafkha, and R.~Salvador, ``Attacking at non-harmonic frequencies in screaming-channel attacks,'' in \emph{International Conference on Smart Card Research and Advanced Applications}.\hskip 1em plus 0.5em minus 0.4em\relax Springer, 2023.

\bibitem{agrawal2003multi}
D.~Agrawal, J.~R. Rao, and P.~Rohatgi, ``Multi-channel attacks,'' in \emph{Cryptographic Hardware and Embedded Systems--CHES 2003}.\hskip 1em plus 0.5em minus 0.4em\relax Springer, 2003.

\bibitem{standaert2008using}
F.-X. Standaert and C.~Archambeau, ``Using subspace-based template attacks to compare and combine power and electromagnetic information leakages,'' in \emph{International Workshop on Cryptographic Hardware and Embedded Systems}.\hskip 1em plus 0.5em minus 0.4em\relax Springer, 2008.

\bibitem{elaabid2011combined}
M.~A. Elaabid, O.~Meynard, S.~Guilley, and J.-L. Danger, ``Combined side-channel attacks,'' in \emph{Information Security Applications: 11th International Workshop, WISA 2010}.\hskip 1em plus 0.5em minus 0.4em\relax Springer, 2011.

\bibitem{hutter2012exploiting}
M.~Hutter, M.~Kirschbaum, T.~Plos, J.-M. Schmidt, and S.~Mangard, ``Exploiting the difference of side-channel leakages,'' in \emph{Constructive Side-Channel Analysis and Secure Design: Third International Workshop, COSADE}.\hskip 1em plus 0.5em minus 0.4em\relax Springer, 2012.

\bibitem{souissi2012towards}
Y.~Souissi, S.~Bhasin, S.~Guilley, M.~Nassar, and J.-L. Danger, ``Towards different flavors of combined side channel attacks,'' in \emph{Topics in Cryptology--CT-RSA 2012: The Cryptographers’ Track at the RSA Conference}.\hskip 1em plus 0.5em minus 0.4em\relax Springer, 2012, pp. 245--259.

\bibitem{heyszl2014clustering}
J.~Heyszl, A.~Ibing, S.~Mangard, F.~De~Santis, and G.~Sigl, ``Clustering algorithms for non-profiled single-execution attacks on exponentiations,'' in \emph{Smart Card Research and Advanced Applications: 12th International Conference, CARDIS}.\hskip 1em plus 0.5em minus 0.4em\relax Springer, 2014.

\bibitem{specht2015improving}
R.~Specht, J.~Heyszl, M.~Kleinsteuber, and G.~Sigl, ``Improving non-profiled attacks on exponentiations based on clustering and extracting leakage from multi-channel high-resolution em measurements,'' in \emph{International Workshop on Constructive Side-Channel Analysis and Secure Design}.\hskip 1em plus 0.5em minus 0.4em\relax Springer, 2015.

\bibitem{yang2017multi}
W.~Yang, Y.~Zhou, Y.~Cao, H.~Zhang, Q.~Zhang, and H.~Wang, ``Multi-channel fusion attacks,'' \emph{IEEE Transactions on Information Forensics and Security}, vol.~12, no.~8, 2017.

\bibitem{genevey2019combining}
C.~Genevey-Metat, B.~Gérard, and A.~Heuser, ``Combining sources of side-channel information,'' in \emph{C\&ESAR 2019}, 2019.

\bibitem{yang2023mca}
W.~Yang, X.~Xiang, C.~Huang, A.~Fu, and Y.~Yang, ``Mca-based multi-channel fusion attacks against cryptographic implementations,'' \emph{IEEE Journal on Emerging and Selected Topics in Circuits and Systems}, 2023.

\bibitem{ayoub2025phasesca}
P.~Ayoub, A.~Hernandez, R.~Cayre, A.~Francillon, and C.~Maurice, ``Phasesca: Exploiting phase-modulated emanations in side channels,'' \emph{IACR Transactions on Cryptographic Hardware and Embedded Systems}, vol. 2025, no.~1, 2025.

\bibitem{kurita2019principal}
T.~Kurita, ``Principal component analysis (pca),'' \emph{Computer Vision: A Reference Guide}, 2019.

\bibitem{specht2018dividing}
R.~Specht, V.~Immler, F.~Unterstein, J.~Heyszl, and G.~Sig, ``Dividing the threshold: Multi-probe localized em analysis on threshold implementations,'' in \emph{IEEE International Symposium on Hardware Oriented Security and Trust (HOST)}.\hskip 1em plus 0.5em minus 0.4em\relax IEEE, 2018.

\bibitem{hettwer2020deep}
B.~Hettwer, D.~Fennes, S.~Leger, J.~Richter-Brockmann, S.~Gehrer, and T.~G{\"u}neysu, ``Deep learning multi-channel fusion attack against side-channel protected hardware,'' in \emph{57th ACM/IEEE Design Automation Conference (DAC)}.\hskip 1em plus 0.5em minus 0.4em\relax IEEE, 2020.

\bibitem{xanthopoulos2013linear}
P.~Xanthopoulos, P.~M. Pardalos, T.~B. Trafalis, P.~Xanthopoulos, P.~M. Pardalos, and T.~B. Trafalis, ``Linear discriminant analysis,'' \emph{Robust data mining}, 2013.

\bibitem{camurati2020understanding}
G.~Camurati, A.~Francillon, and F.-X. Standaert, ``Understanding screaming channels: From a detailed analysis to improved attacks,'' \emph{IACR transactions on cryptographic hardware and embedded systems}, no.~3, 2020.

\bibitem{durvaux2016improved}
F.~Durvaux and F.-X. Standaert, ``From improved leakage detection to the detection of points of interests in leakage traces,'' in \emph{Advances in Cryptology--EUROCRYPT 2016}.\hskip 1em plus 0.5em minus 0.4em\relax Springer, 2016.

\bibitem{poussier2016simple}
R.~Poussier, F.-X. Standaert, and V.~Grosso, ``Simple key enumeration (and rank estimation) using histograms: An integrated approach,'' in \emph{Cryptographic Hardware and Embedded Systems--CHES}.\hskip 1em plus 0.5em minus 0.4em\relax Springer, 2016.

\bibitem{kopf2007information}
B.~K{\"o}pf and D.~Basin, ``An information-theoretic model for adaptive side-channel attacks,'' in \emph{Proceedings of the 14th ACM conference on Computer and communications security}, 2007.

\bibitem{wang2020far}
R.~Wang, H.~Wang, and E.~Dubrova, ``Far field em side-channel attack on aes using deep learning,'' in \emph{Proceedings of the 4th ACM Workshop on Attacks and Solutions in Hardware Security}, 2020.

\bibitem{wang2021advanced}
R.~Wang, H.~Wang, E.~Dubrova, and M.~Brisfors, ``Advanced far field em side-channel attack on aes,'' in \emph{Proceedings of the 7th ACM on Cyber-Physical System Security Workshop}, 2021.

\end{thebibliography}

\end{document}